\documentclass[cameraready]{Interspeech}
\makeatletter
\newcommand{\APEX}{APEX}
\makeatother

\title{\APEX{}: Audio Prototype EXplanations for Classification Tasks}

\author[affiliation={1,2}, orcid=0000-0002-2025-0547, equalcontribution]{Piotr}{Kawa}
\author[affiliation={3,4}, orcid=0000-0002-4085-935X, equalcontribution]{Kornel}{Howil}
\author[affiliation={3,4,5}, orcid=0000-0002-5715-4428]{Piotr}{Borycki}
\author[affiliation={4}]{Miłosz}{Adamczyk}
\author[affiliation={3,4}, orcid=0000-0003-0097-5521]{\\Przemysław}{Spurek}
\author[affiliation={1}, orcid=0000-0002-0266-5802]{Piotr}{Syga}

\address{
    $^1$ Department of Artificial Intelligence, Wroclaw University of Science and Technology, Poland \\
    $^2$ Resemble AI, USA \\
    $^3$ IDEAS Research Institute, Poland \\
    $^4$ Faculty of Mathematics and Computer Science, Jagiellonian University, Poland \\
    $^5$ Doctoral
School of Exact and Natural Sciences, Jagiellonian University, Poland
}

\email{piotr.kawa@pwr.edu.pl, kornel.howil@student.uj.edu.pl}


\usepackage{comment}
\usepackage{amssymb}
\usepackage{amsmath}
\usepackage{tabularx}
\usepackage{multirow}
\usepackage{graphicx}
\usepackage{subcaption}
\usepackage[T1]{fontenc}

\begin{document}

\maketitle

\begin{abstract}

Explainable AI (XAI) has achieved remarkable success in image classification, yet the audio domain lacks equally mature solutions. Current methods apply vision-based attribution techniques to spectrograms, overlooking fundamental differences between visual and acoustic signals. While prototype reasoning is promising, acoustic similarity remains multidimensional. We introduce \APEX{} (Audio Prototype EXplanations), a post-hoc framework for interpreting pre-trained audio classifiers. Crucially, \APEX{} requires no fine-tuning of the original backbone and strictly preserves output invariance. \APEX{} disentangles explanations into four perspectives: Square-based prototypes to localize transient events, Time-based for temporal patterns, Frequency-based highlighting spectral bands, and Time-Frequency-based integrating both. This yields intuitive, example-based explanations that respect acoustic properties, providing greater semantic clarity than standard gradient-based methods.
\end{abstract}

\section{Introduction}

Neural networks dominate modern audio processing, from deepfake detection to healthcare, often surpassing human performance~\cite{human-perception-df,murmur-detection}. However, their deployment in safety-critical environments raises significant ethical and legal concerns, particularly under regulations like the AI Act, necessitating robust interpretability. Current explanations are typically categorized as either feature-based or example-based, or ante-hoc (trained by design) or post-hoc (applied after training).

Despite progress, audio interpretability faces major limitations. Standard feature attribution methods often fail in consistency and reliability~\cite{shen25b_interspeech}. Conversely, inherently interpretable models like AudioProtoPNet~\cite{audioprotopnet} offer intuitive explanations but require training specialized architectures from scratch, rendering them inapplicable to existing high-performance backbones. Furthermore, LLM-based approaches carry risks of hallucination~\cite{xie2026ftgrpo}. Consequently, stable post-hoc methods capable of providing semantically meaningful explanations grounded in the time-frequency nature of sound remain scarce.

\begin{figure}
    \centering
    \includegraphics[width=1\linewidth]{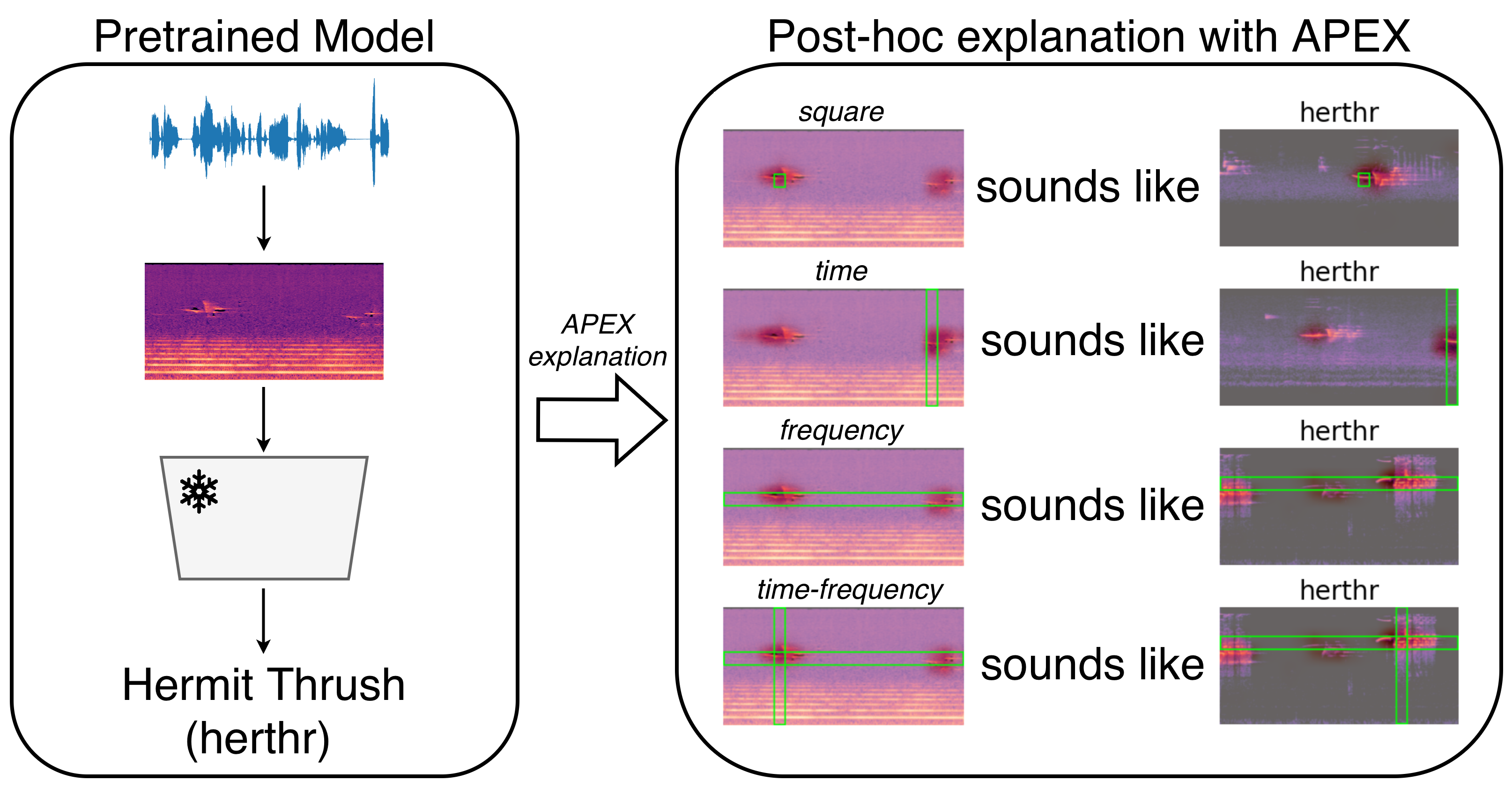}
    \caption{Overview of the \APEX{} framework. Unlike traditional prototype-based approaches that require training specialized architectures from scratch, \APEX{} operates in a post-hoc setting, providing interpretability for arbitrary pre-trained audio backbones. The diagram illustrates our four distinct prototype extraction schemes: Square-based, Time-based, Frequency-based, and Time-Frequency-based. These schemes disentangle the latent space to capture transient events, temporal patterns, spectral characteristics, or hybrid features, respectively.}
    \label{fig:teaser}
    \vspace{-0.3cm}
\end{figure}

In this paper, we introduce \APEX{} (Audio Prototype EXplanations), a novel audio post-hoc interpretability method inspired by prototype-based vision frameworks like \cite{borycki2025epic,struski2024infodisent,dubovik2025side}. In this way, we address the gap between high-performance ''black box'' audio models and human-interpretable concepts. Existing XAI approaches commonly transfer computer vision techniques directly to spectrograms, effectively treating audio as static imagery. This strategy ignores the asymmetry of spectrogram axes: the temporal axis encodes sound evolution, while the frequency axis represents pitch and timbre. Treating these dimensions as equivalent spatial coordinates fails to capture the semantic structure of acoustic events.

To this end, \APEX{} operates directly on latent features from arbitrary pre-trained backbones, rather than relying on generic visual attribution. While these latent spaces encode rich semantic information, they are highly entangled. We therefore introduce a~mechanism that applies structured priors to disentangle the feature space into interpretable components. By enforcing the priors, we extract prototypes under four distinct schemes, each reflecting a specific acoustic perspective (cf.~Fig.~\ref{fig:teaser}).

\begin{figure*}
    \centering
\includegraphics[width=0.90\linewidth]{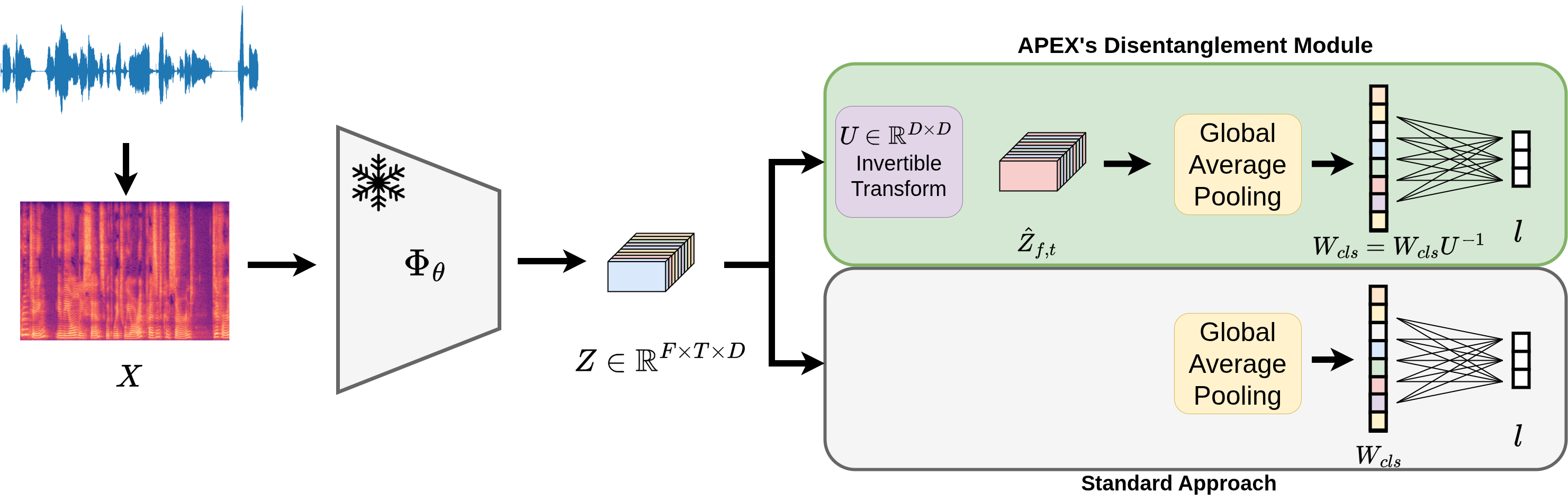}
\includegraphics[width=0.90\linewidth]{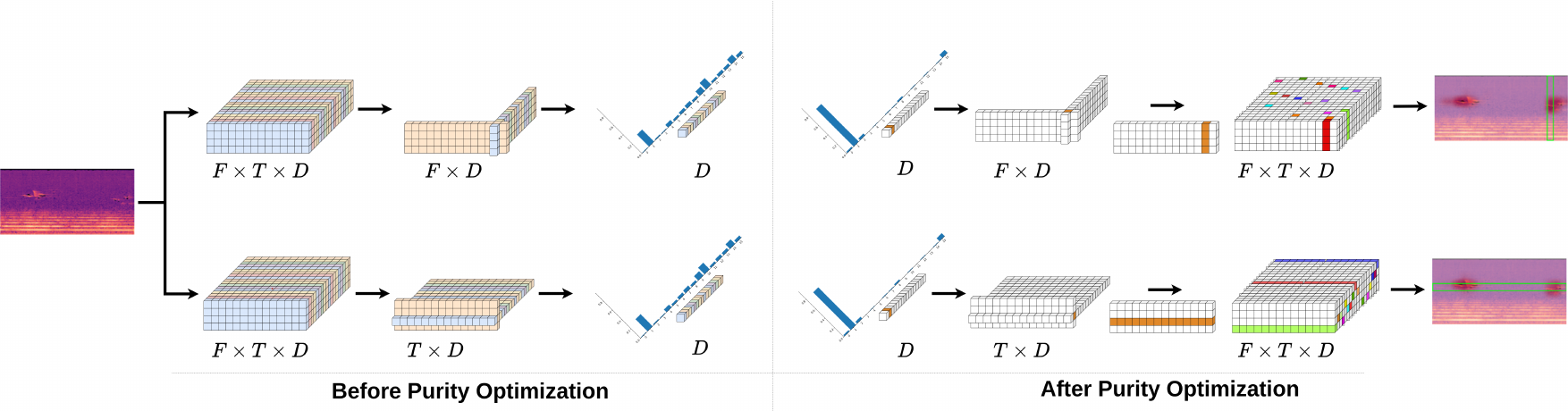}
    \caption{Architectural and representational comparison between a standard audio classifier and the post-hoc APEX framework. \textbf{Top:} While a classical backbone produces entangled feature maps, APEX inserts a Disentanglement Module. Applying a learnable invertible transformation $U$ and its inverse $U^{-1}$ reorganizes the latent space while strictly preserving the original model's predictions (output invariance). \textbf{Bottom:} The effect of purity optimization on the feature space. Before optimization, acoustic concepts are entangled across the channel dimension. Following optimization, the feature maps are disentangled into highly localized, semantically pure prototypes aligned with specific time and frequency components.}
    \label{fig:apex-pipeline}
    \vspace{-0.3cm}
\end{figure*}

First, \textbf{Square-based Prototypes} isolate localized activation peaks in the feature map, capturing transient acoustic events, e.g.,  a bird's chirp or mechanical click, confined to a specific frequency band and duration. Next, \textbf{Time-based Prototypes} aggregate across the frequency axis to identify \textit{when} a defining feature occurs, targeting rhythmic patterns or broadband onsets irrespective of spectral spread. Third, \textbf{Frequency-based Prototypes} aggregate across the time axis to isolate specific frequency bands, explaining decisions driven by time-invariant properties, e.g., a constant engine hum or instrumental timbre, and thus \textit{what} frequency underlies the prediction. Finally, \textbf{Time-Frequency-based Prototypes} combine temporal and spectral components through weighted integration, enabling adaptation to complex events requiring balanced temporal and spectral representation. This structured disentanglement yields explanations grounded in model activations while remaining semantically aligned with human perception.

\noindent The main contributions of this paper are summarized as follows:

\begin{itemize}
    \item We introduce \APEX{}, a novel post-hoc interpretability framework that enables prototype-based reasoning for arbitrary pre-trained audio classifiers. 
    \item We propose a structural disentanglement mechanism that employs four complementary perspectives. By utilizing square-based, time-based, frequency-based, and hybrid weighted schemes, we resolve the ambiguity of acoustic similarity and enable explanations grounded in localized time-frequency events, temporal patterns, or spectral characteristics.
    \item We demonstrate that our multi-view approach provides more precise and semantically meaningful explanations compared to standard gradient-based attribution methods, while maintaining absolute output invariance for the underlying model.
\end{itemize}

\section{Related work}

Neural network interpretability methods are commonly categorized along two orthogonal dimensions: the stage at which interpretation is introduced relative to training, and the type of evidence the explanation is designed to represent. With respect to the training process, methods are typically divided into ante-hoc and post-hoc approaches. Ante-hoc methods integrate interpretability into the model architecture or objective so that explanations arise from the same mechanism that produces predictions. Classical examples include decision trees, while neural network counterparts include prototype-based architectures such as ProtoPNet~\cite{protopnet} and PIP Net~\cite{pipnet}. Post-hoc methods instead explain an already trained model without modifying its parameters or structure. Widely used post-hoc techniques include gradient-based attribution methods such as saliency maps~\cite{saliency-map-example}, Grad-CAM~\cite{gradcam,grad-cam-2}, and Layer-wise Relevance Propagation~\cite{lrp}, as well as perturbation-based, model-agnostic methods such as LIME~\cite{lime} and SHAP~\cite{shap}.

\begin{figure*}
    \centering
    \begin{subfigure}[b]{0.49\textwidth}
        \centering
        Before \APEX{} optimization\\
        \vspace{5pt}
        \fbox{\includegraphics[width=0.95\linewidth]{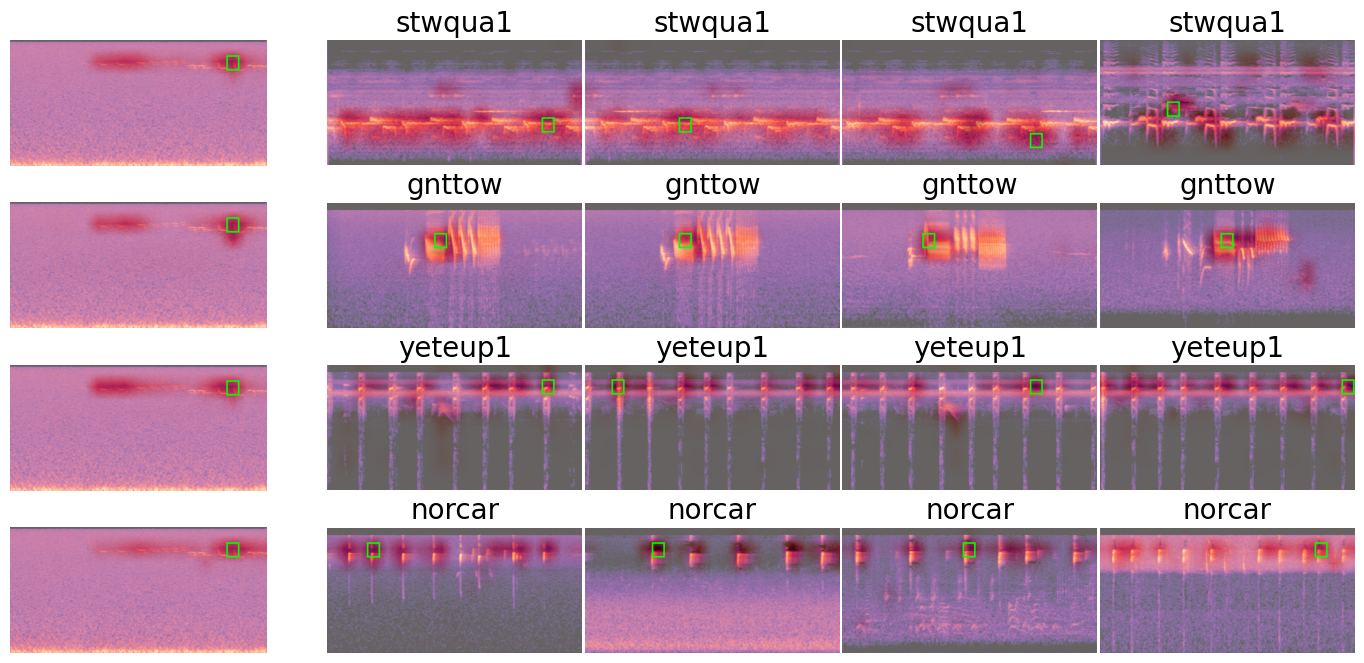}}
    \end{subfigure}
    \begin{subfigure}[b]{0.49\textwidth}
        \centering
        After \APEX{} optimization \\
        \vspace{5pt}
        \fbox{\includegraphics[width=0.95\linewidth]{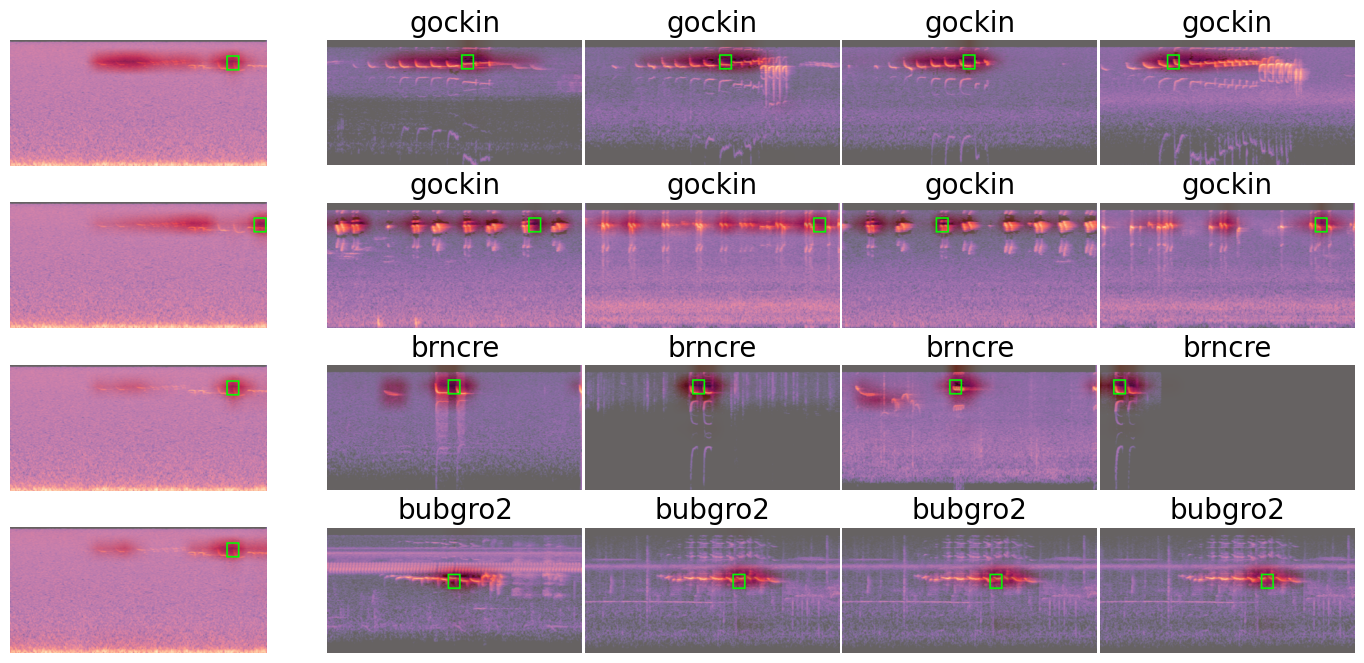}}
    \end{subfigure}
    \caption{Explanations for a Golden-crowned Kinglet (gockin) before and after \APEX{} optimization. Prior to tuning, the extracted prototypes resemble random noise and offer little interpretability. Following \APEX{} optimization, prototypes become semantically meaningful, aligning with the distinct acoustic features present in the input spectrogram. Labels above prototypes show their classes.}
    \label{fig:before_after_apex}
\end{figure*}

Independently of training stage, explanation methods differ in the evidence they provide. Feature attribution methods assign relevance to input components using gradients, gradient activation combinations such as Grad-CAM, or output changes under systematic perturbations as in LIME and SHAP, while example-based methods explain predictions through similarity to representative training instances or learned prototypes.

Compared to vision, audio interpretability is less standardized because time-frequency explanations are not uniquely invertible to waveform or perceptual evidence, and evaluation is complicated by phase, windowing, and non-linear spectral perception, leading to qualitative analysis or task-specific proxies that hinder cross-task and dataset comparability.

Within post-hoc audio attribution,~\cite{gupta24b_interspeech} propose Phoneme Discretized Saliency Maps that discretize saliency by phoneme boundaries for explainable DF detection, though validation is limited to Tacotron2~\cite{tacotron-2} and FastSpeech2~\cite{fastspeech-2}. Similarly,~\cite{channing2024robustrealworldaudiodeepfake} study occlusions and attention visualization for DF interpretability. AudioLIME~\cite{audio-lime} adapts LIME approach to music data by operating on source separation estimates instead of superpixels of the spectrogram images.

Recent inherently interpretable, prototype-based models partially mitigate audio-specific issues, yielding example-level evidence that can be inspected visually or acoustically. AudioProtoPNet~\cite{audioprotopnet} adapts ProtoPNet to multi-label bird classification by learning class-specific prototypes in an embedding space derived from log-Mel-spectrograms and a ConvNeXt-style backbone~\cite{convnext}, enabling classification by similarity to learned prototypes. It scales to a large label space and links predictions to prototypical training patterns, yet it inherits common prototype-learning limitations, including sensitivity to prototype diversity and dataset bias, exhibits limited time-frequency localization, and is restricted to bird-sound classification. SonicProtoPNet~\cite{sonic-protopnet} extends prototypical reasoning to broader audio classification tasks, including environmental sounds, music genre classification, and instrument recognition, and introduces back-soundification to generate playable prototype audio snippets for auditory inspection~\cite{sonic-protopnet}. This improves usability but remains constrained by the chosen front-end representation and reconstruction process, and it does not fully address faithfulness or robustness under common audio transformations.

A complementary line of work uses LLM-based explanation interfaces for audio. The method in~\cite{xie2026ftgrpo} presents an interpretable all-type audio deepfake detector built on an audio LLM, where explanations are frequency- and time-grounded rationales optimized via a constrained reinforcement learning strategy. The approach targets generalization across heterogeneous audio types and explicitly addresses that label-only fine-tuning can yield opaque decision processes, while unconstrained reinforcement learning can induce reward hacking or hallucinated explanations. Similarly, Wu et al.~\cite{AND} propose a neuron-level interpretability framework for deep acoustic models that retrieves highly activating audio examples for each neuron and uses an LLM to generate open-vocabulary natural-language descriptions of neuron response patterns. 

Grinberg et al.~\cite{grinberg25_interspeech} propose a diffusion-based explainability framework for audio deepfake detection that directly learns time-frequency artifact localization using supervision derived from a pair of task-specific real and vocoded speech. Finally,~\cite{shen25b_interspeech} evaluates the reliability of standard feature attribution methods for speech classification through agreement across random seeds, finding generally low attribution consistency even when classifiers agree on predictions. They further show a strong dependence on input representation and attribution granularity, motivating speech-specific attribution design and evaluation.

\section{\APEX: Audio Prototype EXplanations for Classification Tasks}
\label{sec:methodology}

In this section, we describe the proposed \APEX{} framework (see Fig.~\ref{fig:apex-pipeline}). \APEX{} is a post-hoc interpretability method designed for pre-trained audio classification networks. It provides prototype-based explanations by identifying training samples semantically similar to a given query and highlighting relevant time frames, frequency bands, or time-frequency regions. Crucially, our approach disentangles the latent feature space, aligning specific channels with distinct acoustic concepts without degrading the original model's performance.

\begin{figure*}
    \centering
    \setlength{\fboxsep}{0pt} 

    \begin{subfigure}[b]{0.24\textwidth}
        \centering
        Input Spectrogram \\
        \vspace{5pt}
        \fbox{\includegraphics[width=\dimexpr\linewidth-2\fboxrule\relax]{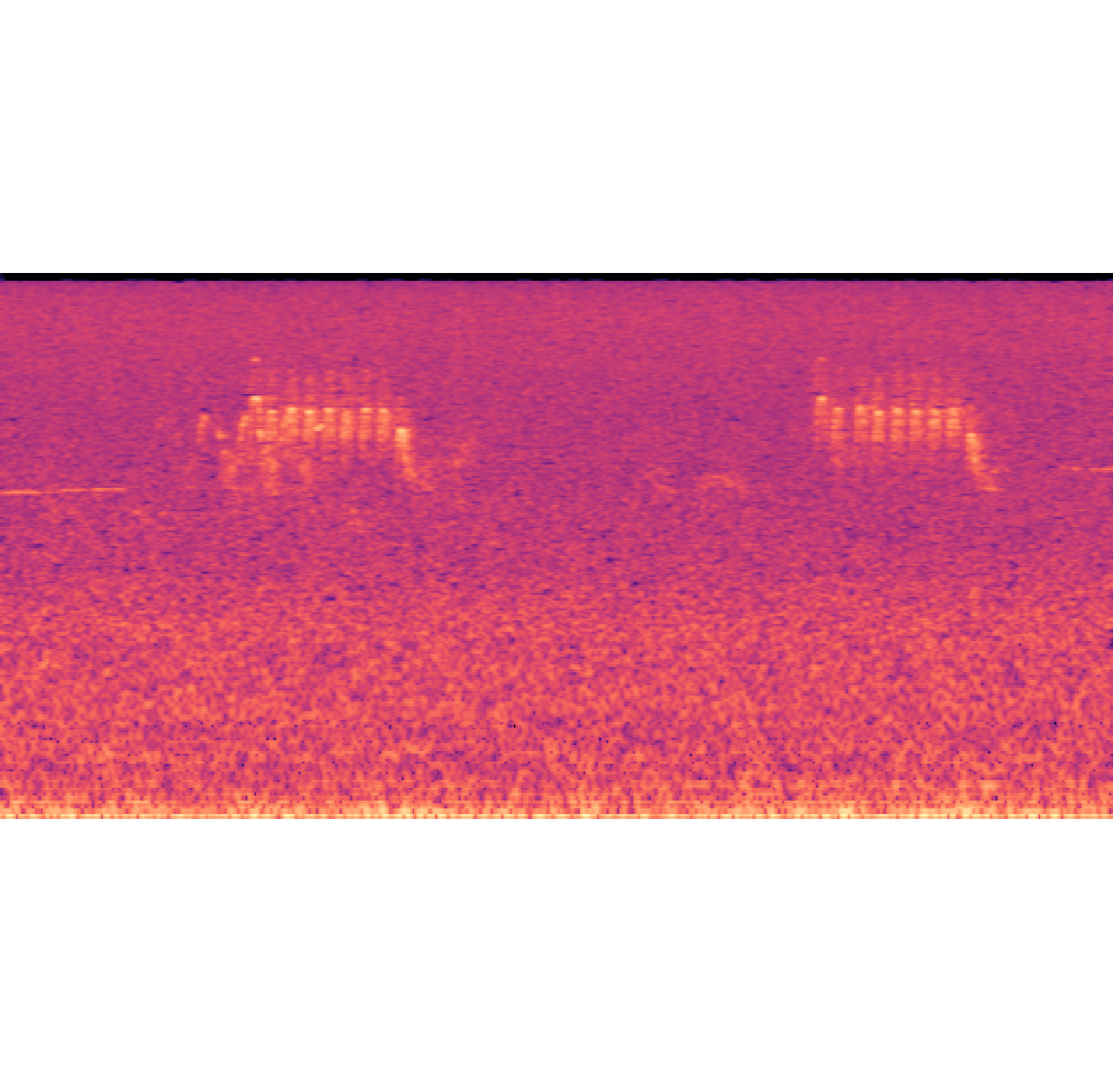}}
    \end{subfigure}\hfill
    \begin{subfigure}[b]{0.24\textwidth}
        \centering
        \APEX{} (our) \\
        \vspace{5pt}
        \fbox{\includegraphics[width=\dimexpr\linewidth-2\fboxrule\relax]{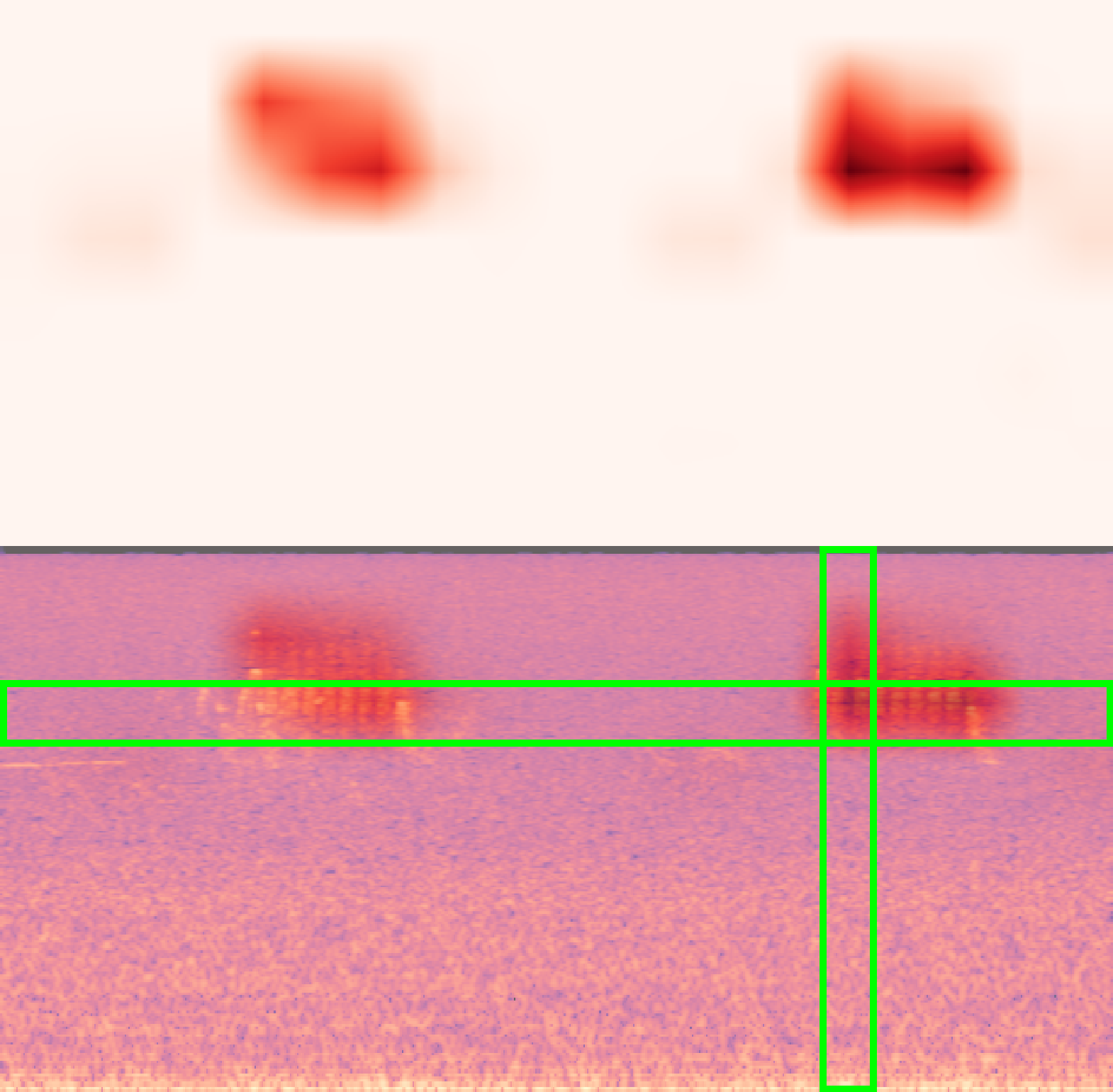}}
    \end{subfigure}\hfill
    \begin{subfigure}[b]{0.24\textwidth}
        \centering
        Grad-CAM \\
        \vspace{5pt}
        \fbox{\includegraphics[width=\dimexpr\linewidth-2\fboxrule\relax]{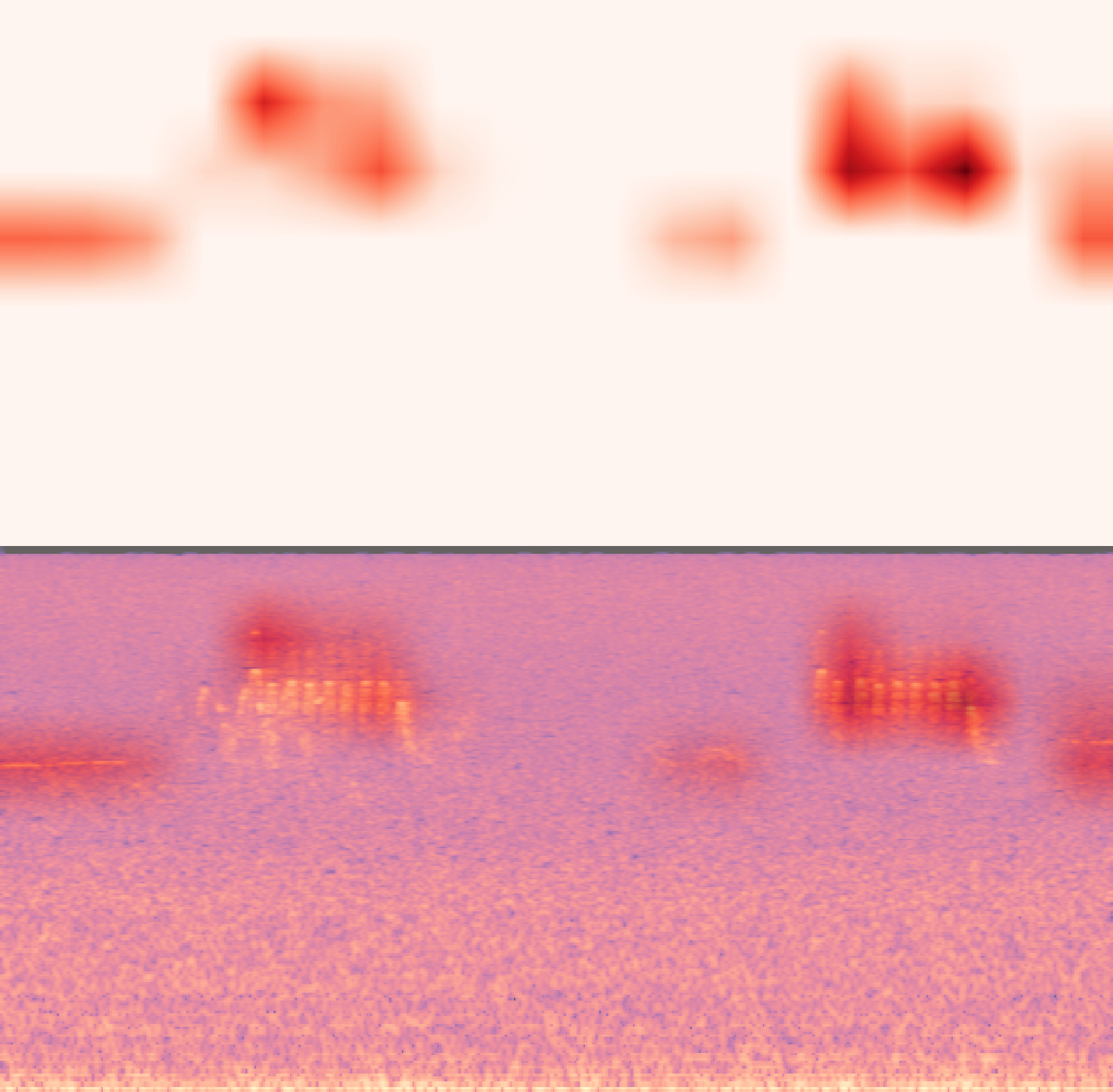}}
    \end{subfigure}\hfill
    \begin{subfigure}[b]{0.24\textwidth}
        \centering
        LIME \\
        \vspace{5pt}
        \fbox{\includegraphics[width=\dimexpr\linewidth-2\fboxrule\relax]{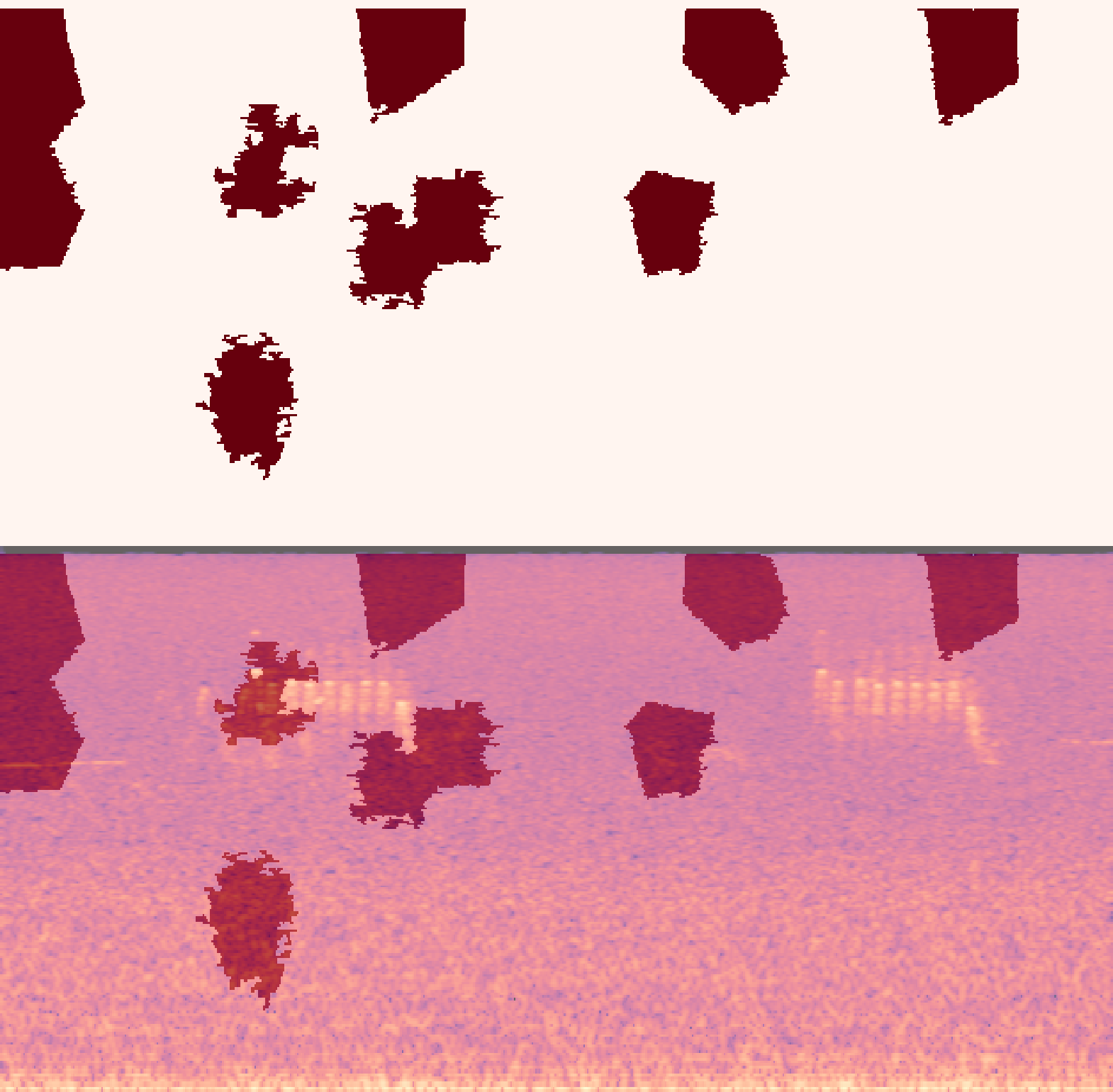}}
    \end{subfigure}

    \vspace{1pt} 

    \begin{subfigure}[b]{0.24\textwidth}
        \centering
        \fbox{\includegraphics[width=\dimexpr\linewidth-2\fboxrule\relax]{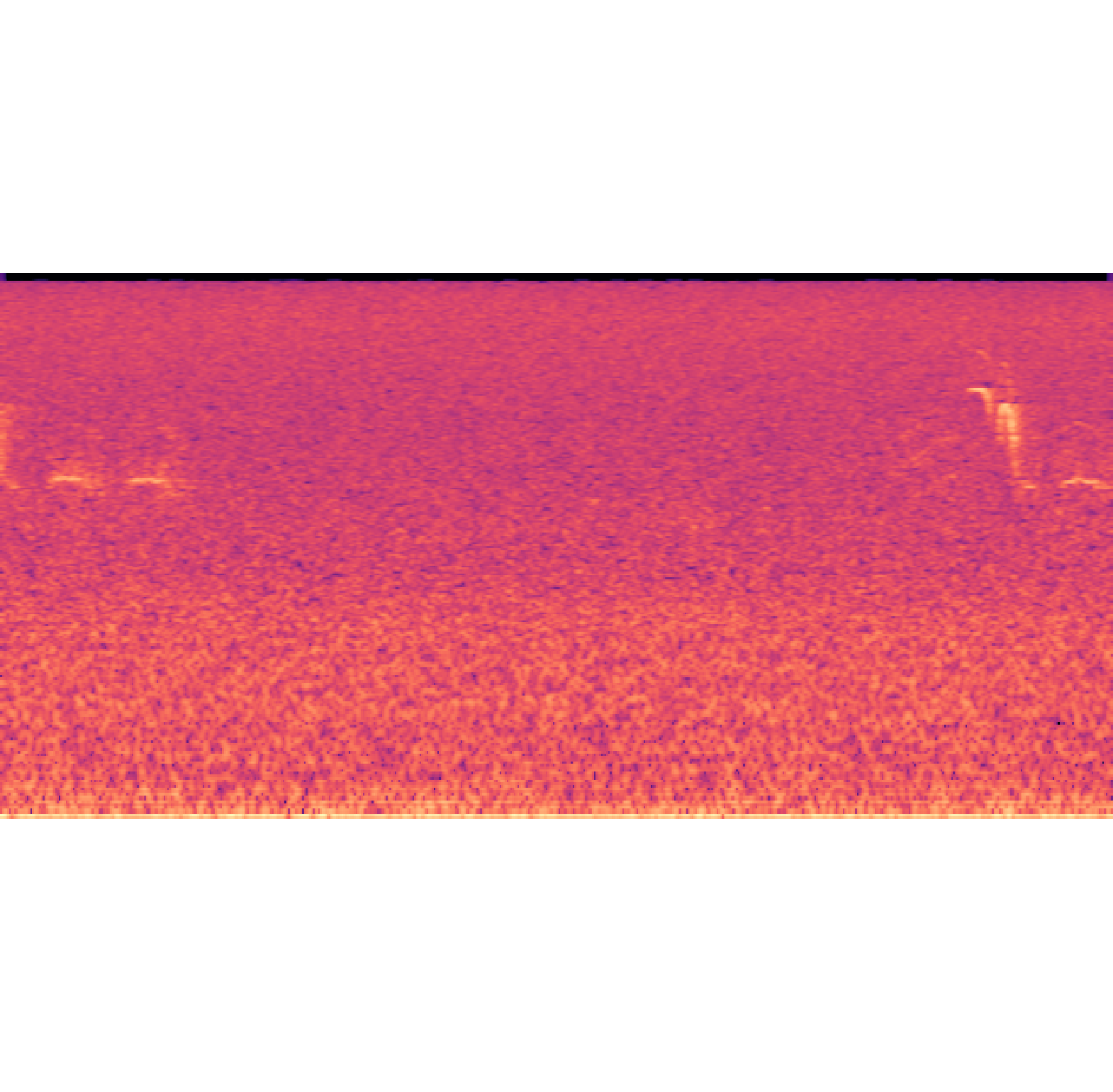}}
    \end{subfigure}\hfill
    \begin{subfigure}[b]{0.24\textwidth}
        \centering
        \fbox{\includegraphics[width=\dimexpr\linewidth-2\fboxrule\relax]{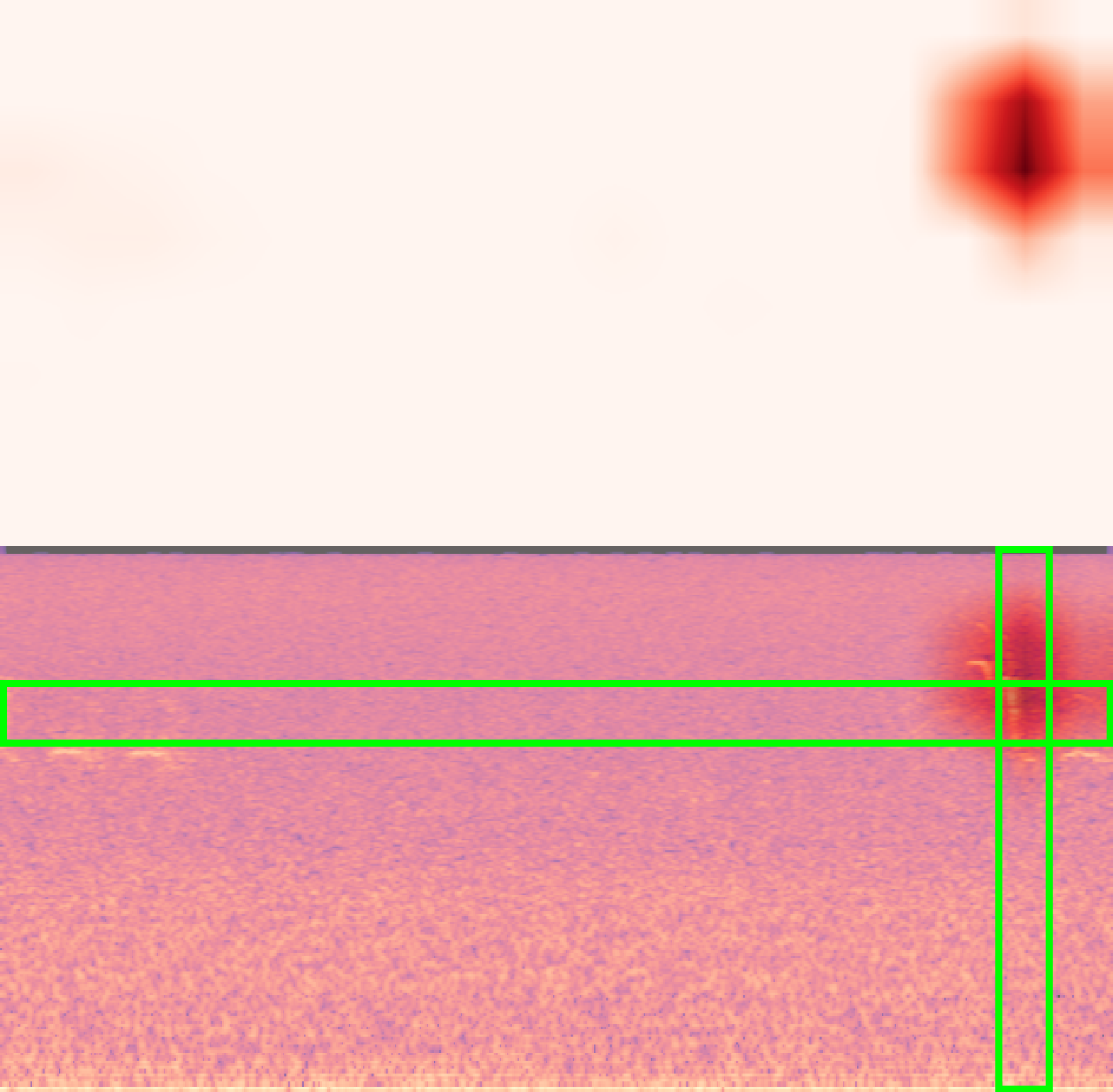}}
    \end{subfigure}\hfill
    \begin{subfigure}[b]{0.24\textwidth}
        \centering
        \fbox{\includegraphics[width=\dimexpr\linewidth-2\fboxrule\relax]{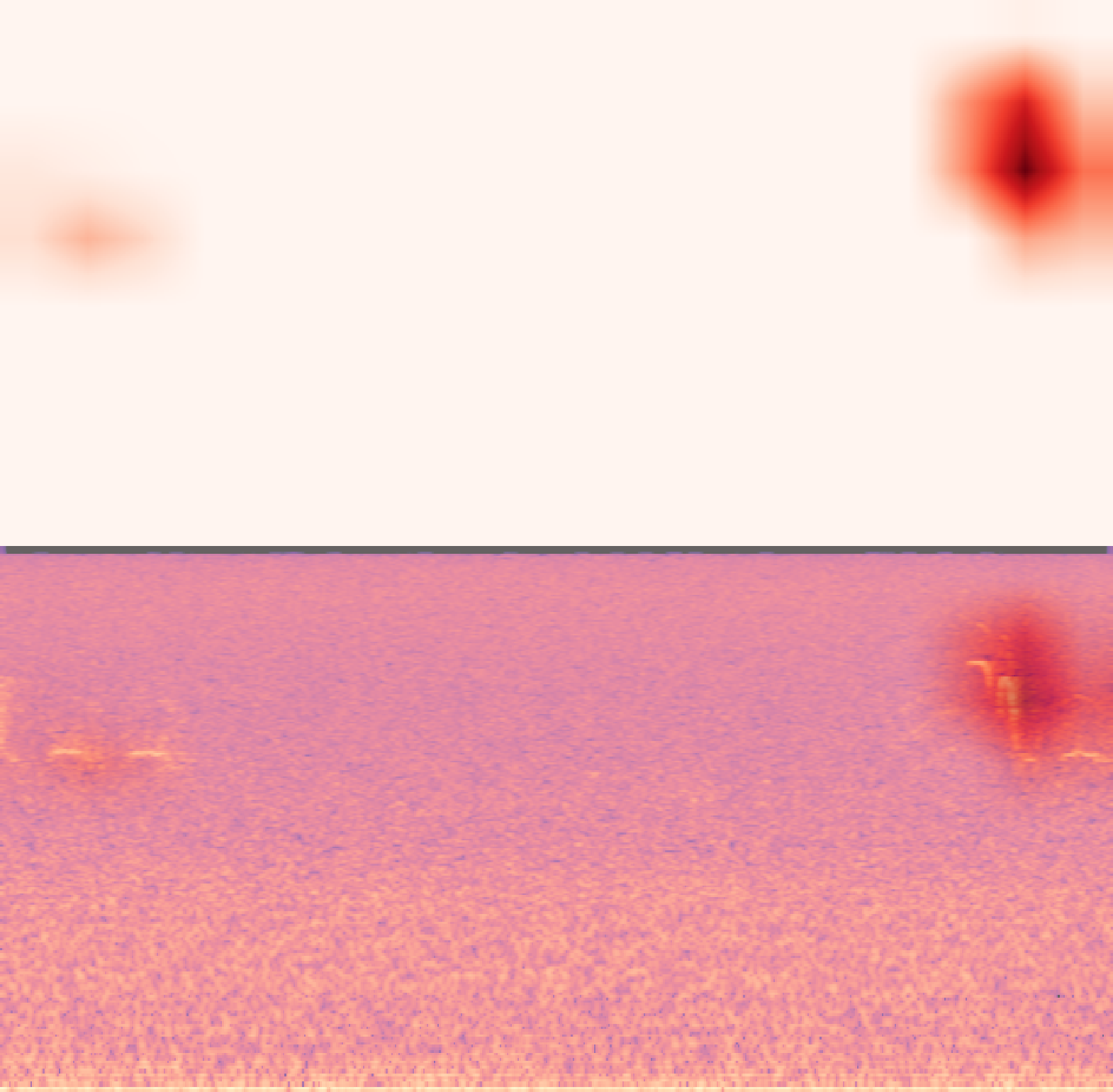}}
    \end{subfigure}\hfill
    \begin{subfigure}[b]{0.24\textwidth}
        \centering
        \fbox{\includegraphics[width=\dimexpr\linewidth-2\fboxrule\relax]{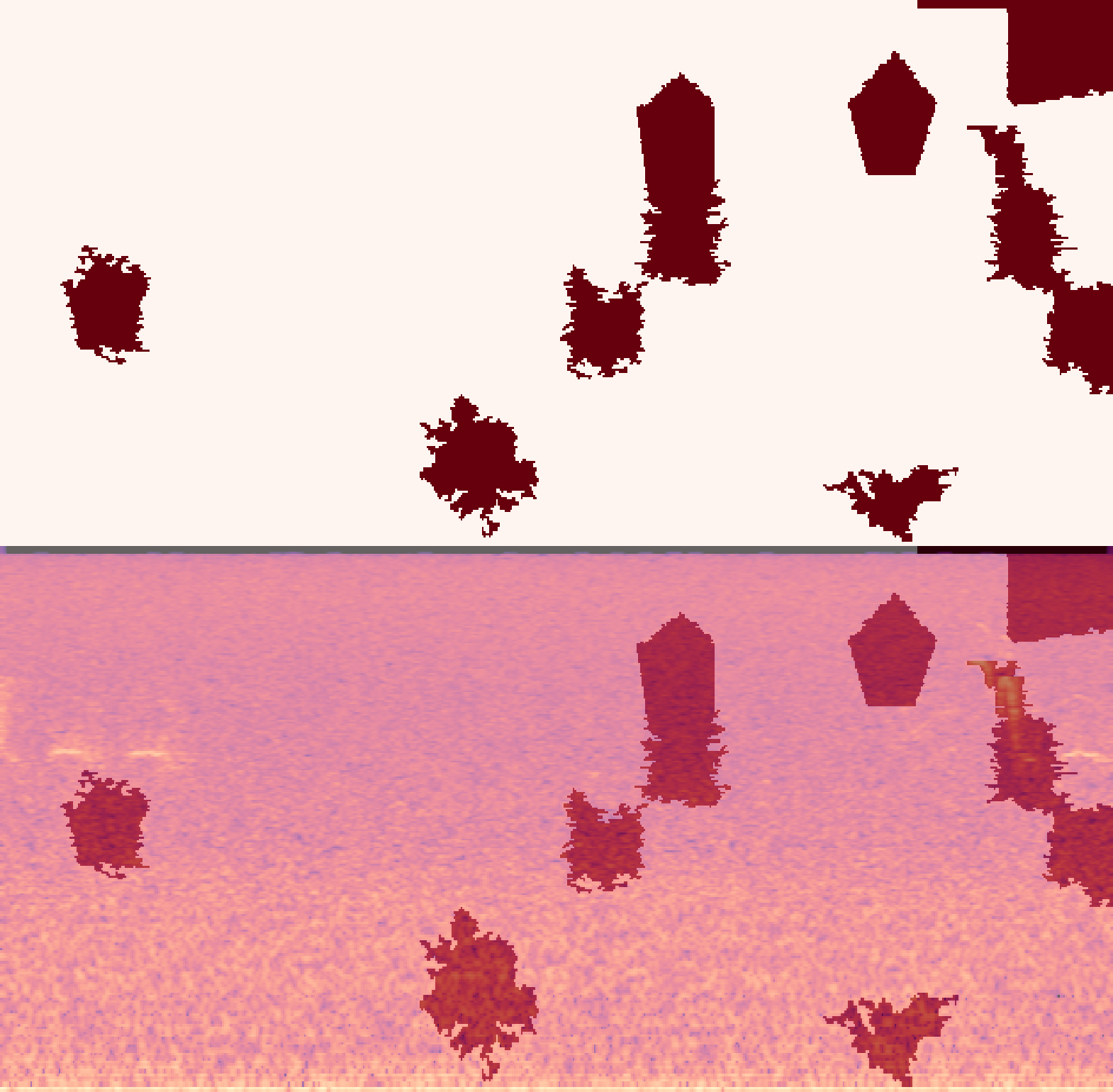}}
    \end{subfigure}

    \caption{Qualitative comparison of post-hoc interpretability methods conducted on a representation learned on top of a pretrained ConvNeXt classifier. The input spectrogram (far left) contains distinct correctly classified test sounds from the SNE test set of the BirdSet \cite{birdset} dataset. \APEX{} framework successfully disentangles the latent space to generate highly localized, semantically clear time-frequency explanations. Furthermore, \APEX{} grounds these regions in representative acoustic prototypes from the training data; a capability entirely missing from standard attribution methods. In contrast, standard vision-based attribution methods like Grad-CAM produce diffuse, unconstrained heatmaps, while LIME yields fragmented regions that fail to accurately capture the acoustic boundaries.}
    \label{fig:posthoc}
    \vspace{-0.3cm}
\end{figure*}

\textbf{Preliminaries and Problem Setup}
We consider a standard audio classification network $\Phi$ trained on spectral-domain representations (e.g., linear or Mel-spectrograms). Let $X \in \mathbb{R}^{F_{in} \times T_{in}}$ be an input spectrogram.
The network consists of a convolutional backbone $\Phi_{\Theta}$ and a linear classification head.

\noindent The backbone processes the input to produce a latent feature tensor $Z_{f,t}$:
\begin{equation}
    Z_{f,t} = \Phi_{\Theta}(X) \in \mathbb{R}^{F \times T \times D},
\end{equation}
where $F$ and $T$ denote the reduced frequency and time dimensions of the feature map, and $D$ denotes the number of channels. In standard architectures, this feature map is aggregated via Global Average Pooling (GAP) to yield a fixed-length feature vector 
$\mathbf{v} = \mathrm{GAP}(Z) \in \mathbb{R}^{D}$.
This vector is subsequently projected by a linear layer with weights $W_{\text{cls}} \in \mathbb{R}^{N \times D}$ to produce class logits $\mathbf{l} \in \mathbb{R}^{N}$ with $N$ being the number of classes.

\textbf{Feature Space Disentanglement}
Standard training procedures often lead to entangled representations, in which a single acoustic concept is distributed across multiple channels. To address this, we introduce a \textbf{Disentanglement Module} inserted between the backbone and the pooling layer. This module applies a learnable, invertible linear transformation, parameterized by a square matrix $U \in \mathbb{R}^{D \times D}$. The disentangled feature map $\hat{Z}_{f,t}$ is obtained as:
\begin{equation}
    \hat{Z}_{f,t} = (U \cdot Z_{f,t}) \in \mathbb{R}^D.
\end{equation}
To guarantee the invertibility of $U$, while allowing its optimization, we parameterize it as the matrix exponential of a square matrix $A \in \mathbb{R}^{D \times D}$: $U = \exp(A)$. Due to properties of the matrix exponential, the obtained matrix $U$ is always invertible, with its inverse $U^{-1}=\exp(-A)$. This follows from standard results on the matrix exponential, which maps any square matrix to an element of the general linear group $\mathrm{GL}_D(\mathbb{R})$~\cite{MatrixAnalysis}. 

The goal of $U$ is to re-express the feature space in a basis where channels are independent and semantically pure. In particular, the transformation is optimized to disentangle the learned representation according to the purity measure introduced later, encouraging each channel of $\hat{Z}_{f,t}$ to capture a single acoustic factor rather than a mixture of concepts.  

\textbf{Output Invariance.}
A key requirement of our post-hoc approach is to preserve the pre-trained network's original behavior. To guarantee that the model outputs remain unchanged, we apply the inverse transformation $U^{-1}$ to the classification head:
\begin{align}
    \mathbf{v}_{\text{new}} &= \mathrm{GAP}(\hat{Z}) = \mathrm{GAP}(U Z) = U \cdot \mathrm{GAP}(Z), \\
    \mathbf{l}_{\text{new}} &= (W_{\text{cls}} U^{-1}) \cdot \mathbf{v}_{\text{new}} = W_{\text{cls}} U^{-1} U \cdot \mathrm{GAP}(Z) = \mathbf{l}_{\text{old}}.
\end{align}
This ensures that $\mathbf{l}_{\text{new}} = \mathbf{l}_{\text{old}}$, preserving the exact accuracy of the original model while enabling interpretability in the intermediate $\hat{Z}$ space.

\textbf{Audio Prototype Extraction Schemes}
We define a prototype for channel $k$ as a representative feature vector extracted from specific frequency and time coordinates. Unlike visual objects, acoustic concepts can be localized in time, frequency, or both. To capture this, we propose four distinct schemes for identifying the optimal coordinates for a given channel and directly extracting the corresponding feature vector:

\begin{enumerate}
    \item \textbf{Square-based prototypes}: extracts the feature vector from the ``pixel'' with the largest value within a given feature map:
    \begin{gather}
        (f^*, t^*) = \operatorname*{arg\,max}_{f,t} \hat{Z}[f, t, k] \\
        p_{s} = \hat{Z}[f^*, t^*, :]
    \end{gather}

    \item \textbf{Time-based prototypes}: first averages rows of the $k$-th channel (frequencies) and later selects the column (time segment) with the maximum average value:
    \begin{gather}
        t^* = \operatorname*{arg\,max}_{t} \left( \frac{1}{F}\sum_{f=1}^{F}\hat{Z}[f,t,k]\right) \\
        p_{t} =\frac{1}{F}\sum_{f=1}^{F}\hat{Z}[f,t^*,:]
    \end{gather}

    \item \textbf{Frequency-based prototypes}: first averages columns of the $k$-th channel (time frames) and later selects the row (frequency band) with the maximum average value:
    \begin{gather}
        f^* = \operatorname*{arg\,max}_{f} \left( \frac{1}{T}\sum_{t=1}^{T}\hat{Z}[f,t,k]\right) \\
        p_{f} = \frac{1}{T}\sum_{t=1}^{T}\hat{Z}[f^*,t,:]
    \end{gather}

    \item \textbf{Time-Frequency-based prototypes}: an average of time- and frequency-based prototype vectors:
    \begin{equation}
        p_{tf}= \frac{1}{2}(p_{t} +  p_{f}).
    \end{equation}
\end{enumerate}

\textbf{Optimization via Scheme-Dependent Purity}
The objective of the disentanglement process is to maximize the ``purity'' of the prototypes by optimizing the transformation matrix $U$. Ideally, if channel $k$ represents a distinct acoustic concept, the representative feature vector $\mathbf{p}^{(k)}\in\{p_{s}, p_{t}, p_{f}, p_{tf}\}$, extracted from the disentangled map $\hat{Z}$ depending on the scheme, should concentrate its energy solely in channel $k$.

The adaptation of the training process to each of the 4 methods is implemented through the coordinate selection logic, which imposes distinct structural priors on the latent space:

\begin{figure}
    \centering
    \begin{subfigure}[b]{0.47\textwidth}
        \includegraphics[width=0.99\linewidth]{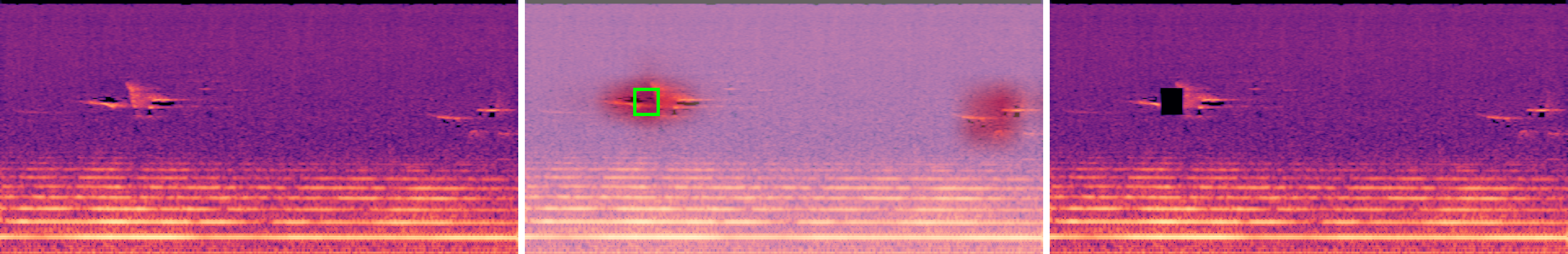}
        \caption{Square}
        \vspace{0.1cm}
    \end{subfigure}
    \begin{subfigure}[b]{0.47\textwidth}
        \includegraphics[width=0.99\linewidth]{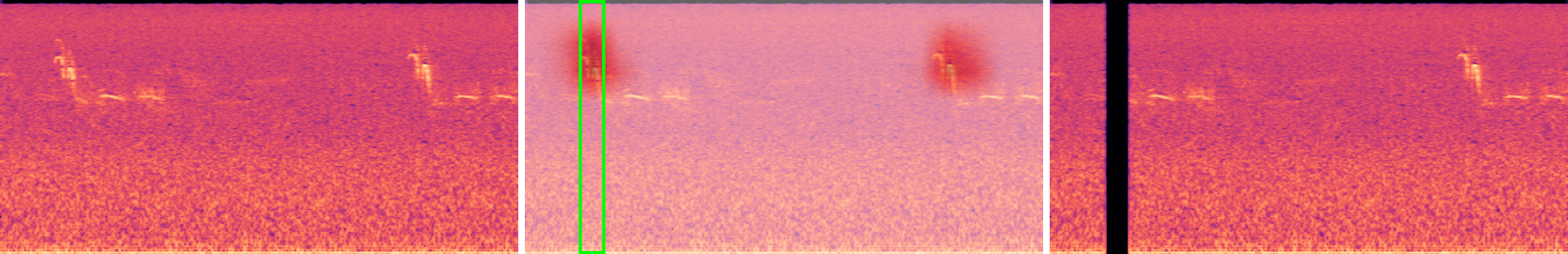}
        \caption{Time}
        \vspace{0.1cm}
    \end{subfigure}
    \begin{subfigure}[b]{0.47\textwidth}
        \includegraphics[width=0.99\textwidth]{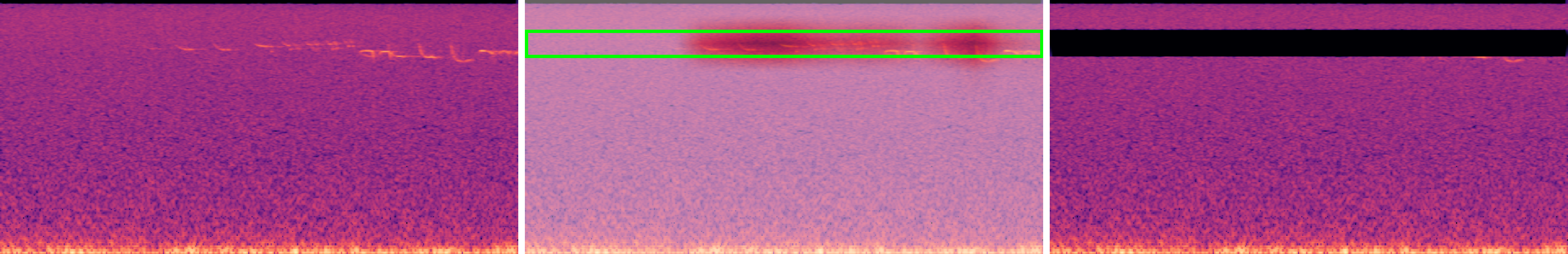}
        \caption{Frequency}
        \vspace{0.1cm}
    \end{subfigure}
    \begin{subfigure}[b]{0.47\textwidth}
        \includegraphics[width=0.99\textwidth]{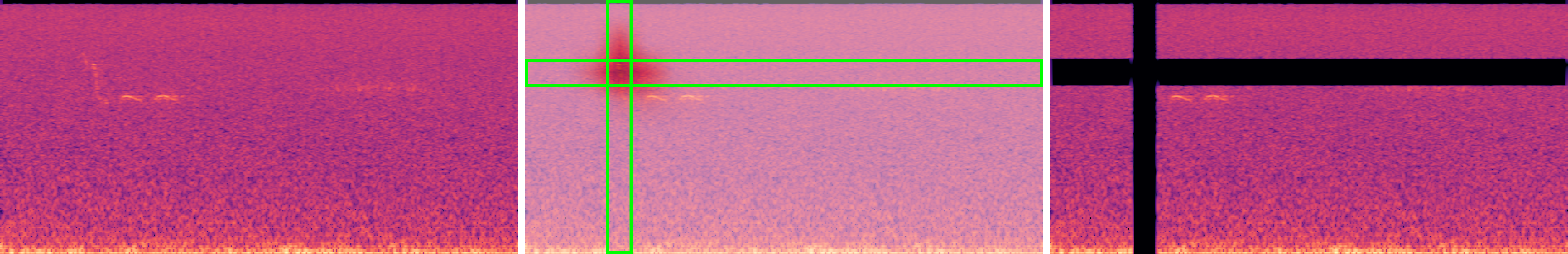}
        \caption{Time-frequency}
    \end{subfigure}
    \caption{Depiction of the \APEX{} masking strategy used to evaluate feature importance. The columns show the original spectrogram (left), the \APEX{} explanation with the localized prototype region highlighted in green (middle), and the corresponding masked spectrogram (right) for each of the schemes.}
    \label{fig:masking}
    \vspace{-0.3cm}
\end{figure}

\begin{enumerate}
    \item {\em Square-based Adaptation:} The model searches for the single most active ``unit'' in the feature map $p_s$. During optimization, $U$ is forced to concentrate the semantics of channel $k$ into the point of highest individual intensity. This promotes the discovery of highly localized, transient acoustic events such as clicks or percussive onsets.
    
\item {\em Time-based Adaptation:} By selecting $t^*$ by frequency-averaging ($p_t$), the optimization becomes invariant to spectral shifts. This encourages the disentanglement module to associate channel $k$ with distinct temporal patterns and rhythmic structures independent of their frequency distribution, making it particularly suitable for broadband events that are primarily defined in time.

    \item {\em Frequency-based Adaptation:} Selecting $f^*$ via time-averaging ($p_f$) makes the optimization invariant to the duration or precise timing of the signal. This encourages $U$ to isolate channels representing stable timbral characteristics, constant harmonic structures, or pitch-related features that persist across multiple time frames.
    
    \item {\em Time-frequency Adaptation:} This approach provides a joint regularization for both temporal and spectral localization. By equally weighting the time- and frequency-based components in the hybrid coordinate $p_{tf}$, the optimization inherently balances ``when'' and ``what'' is being disentangled. This establishes a fixed structural prior that concurrently accounts for both temporal evolution and spectral content.
\end{enumerate}

The \textbf{Purity Score} for channel $k$ serves as the primary training objective:
\begin{equation}
    \operatorname{purity}(I,k) = \frac{|\mathbf{p}^{(k)}_k|}{||\mathbf{p}^{(k)}||_2} \in [0, 1].
\end{equation}

In conventionally trained networks, prototypes are often non-descriptive, with activations uniformly distributed across the channel vector, leading to low purity scores. The task of the disentanglement process is to drive this score towards $\operatorname{purity}=1$. This state is achieved when the values of the prototypical vector $\mathbf{p}^{(k)}$ are zero for all indices except $k$, meaning the channel responds exclusively to a single semantic concept.

Crucially, this optimization by $U$ ensures that while the internal representation becomes interpretable and disentangled, the \textbf{output invariance} is maintained. By adapting the explanation logic to the user's area of interest (e.g., focusing on time vs. frequency), \APEX{} is a versatile tool for auditing audio models without sacrificing their original predictive performance.

\textbf{Prototype Selection}
Once the disentanglement matrix $U$ is optimized, our objective is to identify a set of representative training samples (i.e., prototypes) for each channel. This provides the actual example-based evidence used for explanations.

For each channel $k \in \{1, \dots, D\}$, we define its overall activation for a given input spectrogram $X$ as the sum of its values across the spatial dimensions of the disentangled feature map $\hat{Z}$:
\begin{equation}
    \mathrm{activ}(\hat{Z}, k) = \sum_{f=1}^{F} \sum_{t=1}^{T} \hat{Z}[f, t, k].
\end{equation}

Using this metric, we evaluate all samples within the training dataset and score them based on the magnitude of their response in the $k$-th channel. The $m$ training samples that elicit the highest activation scores serve as the positive prototypes for that channel:
\begin{equation}
    \mathcal{P}_k^{+} = \operatorname*{arg\,top\text{-}m}_{X \in \text{Train}} \mathrm{activ}(\hat{Z}^{(X)}, k).
\end{equation}

Similarly, inputs that elicit the strongest negative responses can be designated as negative prototypes. However, due to their more intuitive and informative nature in identifying present acoustic concepts, our experiments primarily focus on positive prototypes. Reiteration of this process across all $D$ channels gives a global prototype bank where each feature channel is consistently linked to its most representative audio segments. 

\begin{figure}
    \centering
\includegraphics[width=\linewidth]{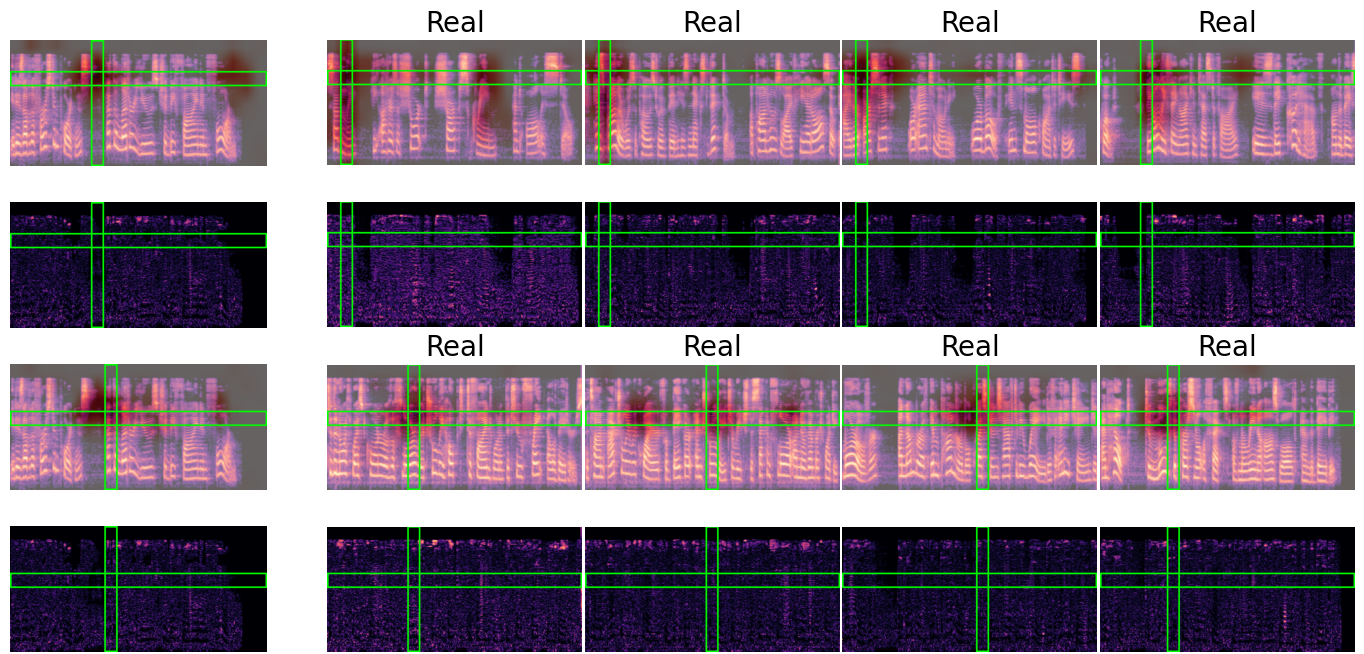}
    \caption{\APEX{} explanations (ConvNeXt) for correctly classified real audio from LJSpeech~\cite{ljspeech}. The left column shows input query spectrograms and heatmaps; the right four show prototypical parts and labels. Odd rows display original spectrograms, while even rows highlight spectral differences between the real and HiFi-GAN vocoded audio. (Trained on the HiFi-GAN subset of WaveFake~\cite{wavefake}).}
    \label{fig:comp_deepfake}
    \vspace{-0.3cm}
\end{figure}

As demonstrated in Fig.~\ref{fig:before_after_apex}, the prototypes extracted from classically trained networks generally suffer from low descriptive power and semantic inconsistency, as predictive concepts remain entangled across multiple channels. Conversely, the prototypes extracted from our optimized $\hat{Z}$ space yield highly coherent, interpretable examples that are closely aligned with distinct acoustic phenomena.

\begin{figure*}
    \centering
    \begin{subfigure}[b]{0.49\textwidth}
        \centering
        \APEX{} (our)\\
        \vspace{5pt}
        \fbox{\includegraphics[width=0.95\linewidth]{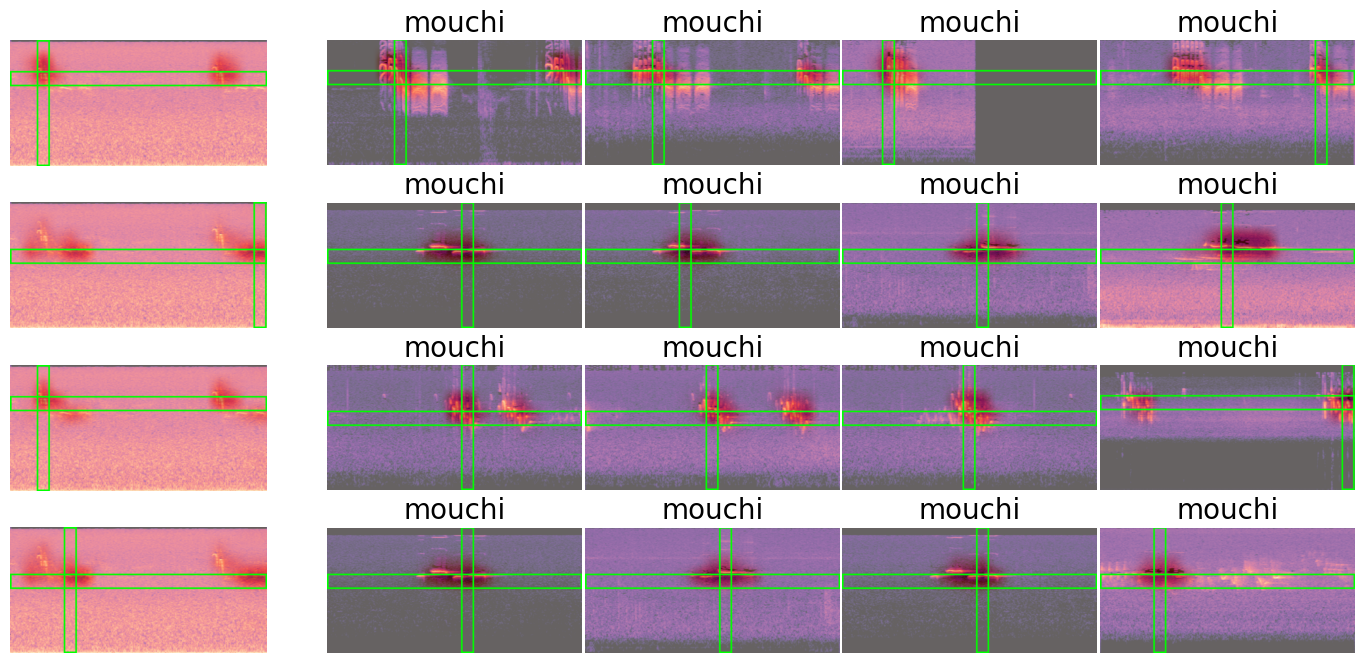}}
        \label{fig:first}
    \end{subfigure}
    \begin{subfigure}[b]{0.49\textwidth}
        \centering
        AudioProtoPNet \\
        \vspace{5pt}
        \fbox{\includegraphics[width=0.95\linewidth]{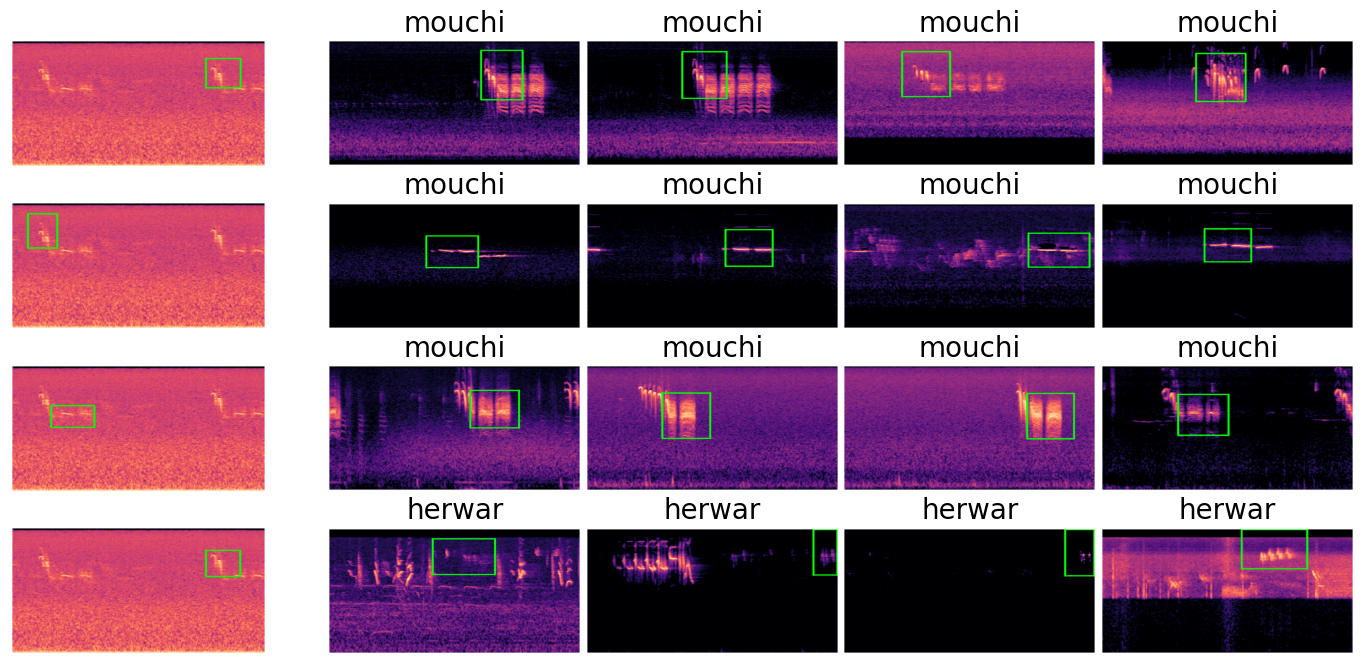}}
        \label{fig:second}
    \end{subfigure}
    \caption{Comparison of explanations between \APEX{}  (our) and prototype-based model AudioProtoPNet. 
    The comparison is conducted on a representation learned on top of a pretrained ConvNeXt. This example shows correctly classified test sound of a mouchi (Mountain Chickadee) from the SNE test set of BirdSet \cite{birdset} dataset. Labels above prototypes show their classes.}
    \label{fig:comp_birdset}
    \vspace{-0.3cm}
\end{figure*}

\textbf{Inference} After training the Disentanglement Module and selecting the channel prototypes, we explain the model's prediction for a given input spectrogram by identifying the top-$k$ channels that contribute most to the predicted class. These contributions are computed by examining the class-specific logits of the final classification layer. Specifically, given an input spectrogram and its predicted class $y$, we analyze the corresponding activations, retaining only positive evidence by applying a ReLU operation prior to evaluating the contributions. In our output explanations, we first highlight the relevant areas of the input spectrograms by overlaying the input with the green horizontal, vertical, or square rectangles depending on the used purity type (see Figures~\ref{fig:apex-pipeline}, ~\ref{fig:before_after_apex}, and ~\ref{fig:posthoc}). 

\textbf{Heatmaps} Inspired by \cite{zhou2016learning}, we additionally extract spatial heatmaps directly from our disentangled feature representations. Unlike standard attribution techniques such as Grad-CAM~\cite{gradcam,grad-cam-2} or LRP~\cite{lrp}, which require a backward pass through the neural network, our method operates solely during the forward pass. Specifically, we compute individual channel heatmaps by scaling the disentangled activations from the final convolutional layer with the classification head's weights (with the inverse transformation $U^{-1}$ applied) corresponding to the predicted class. Rather than summing across channels, each channel retains its distinct spatial activation map, to which a channel-normalized bias term is added. Subsequently, we apply a ReLU activation to discard negative evidence, isolating the regions that strictly contribute positively to the model's decision and constraining the values to a $[0,\,1]$ scale. Finally, the channel-specific heatmaps are upsampled via bilinear interpolation to match the dimensions of the original input spectrogram, yielding highly localized time-frequency explanations for each disentangled prototype. Fig.~\ref{fig:posthoc} illustrates comparison with other post-hoc methods.

\section{Experiments}
We evaluate our approach under two experimental scenarios. In the first scenario, to demonstrate that \APEX{} maintains strict output invariance, we evaluate its classification performance against the vanilla pre-trained ConvNeXt-Base~\cite{convnext} backbone and an AudioProtoPNet trained on the same pre-trained model.

In the second scenario, we investigate whether the spectrogram regions highlighted by \APEX{} are essential for correct classification. First, we identify the latent channel that contributes most strongly to the predicted class. Depending on the prototype extraction scheme, the disentangled feature map of this channel localizes a specific region in the spectrogram. This region corresponds to a square patch (square-based prototypes), a column spanning time (time-based prototypes), a row spanning frequency (frequency-based prototypes), or a joint column–row structure (time–frequency-based prototypes). To assess the importance of these regions, we constrain the features available to the model in a manner similar to prior sub-band ablation studies ~\cite{sriskandaraja16_interspeech,witkowski17_interspeech,mu2021environmental}. Specifically, we apply a soft filter to mask the localized time-frequency regions and re-evaluate the model's performance on the modified input. We perform this evaluation across all four prototype extraction schemes. Fig.~\ref{fig:masking} depicts these masking strategies. For a robust baseline comparison, we also evaluate the model's performance on spectrograms where randomly selected regions of identical dimensions have been masked.

The disentanglement module $U$ is trained for 20 epochs, with prototypes being recalculated every 2 epochs. During this recalculation process, the number of prototypes for each channel is simultaneously reduced; specifically, we begin with 100 prototypes per prototypical channel and linearly decrease this value to 5 by the conclusion of the training stage. For optimization we use Adam \cite{adam} with $lr=10^{-4}$, $\beta_1=0.9$, $\beta_2=0.999$, $\texttt{weight\_decay}=10^{-5}$, and $\texttt{batch\_size}=512$. Optimization time depends on the dataset size and the backbone model. For ConvNeXt-Base with 1024 latent channels,, optimization typically takes a few hours on an NVIDIA A100 GPU.

\subsection{Audio Deepfake Detection}

\textbf{Training} To evaluate our approach, we use the WaveFake dataset~\cite{wavefake}, an audio deepfake detection benchmark derived from the bona fide LJSpeech~\cite{ljspeech} and JSUT~\cite{jsut} corpora and their fake counterparts generated with seven vocoder-based and one text-to-speech system. We restrict our experiments to the LJSpeech subset, comprising 13,100 real utterances, and include samples produced by all available vocoders: HiFi-GAN~\cite{hifi-gan}, WaveGlow~\cite{waveglow}, MelGAN and MelGAN-Large~\cite{melgan}, MultiBand-MelGAN and FullBand-MelGAN~\cite{multiband-melgan}, and Parallel-WaveGAN~\cite{parallel-wavegan}. The TTS-based system is excluded, as intermediate text-to-speech synthesis induces temporal misalignment relative to the original recordings, complicating explanation visualization without increasing task difficulty.

We train ConvNeXt-Base~\cite{convnext} classifiers on the subsets of each deepfake generation method and corresponding bona fide samples. These datasets are partitioned by allocating the first 1,000 real and corresponding 1,000 fake audio samples to the test set, the subsequent 1,000 samples of each class to the validation set, and the remaining data to the training set. Using these baseline ConvNeXt models, we then train AudioProtoPNet \cite{audioprotopnet} and APEX models. AudioProtoPNet \cite{audioprotopnet} is trained according to the default settings described in its original publication.

\begin{table*}[t!]
    \centering
    \caption{Equal Error Rates (EER) and average EER (aEER) expressed in percentages. Rows indicate the specific vocoder dataset used for training, while columns represent the unseen datasets used for testing. Within each cell, results are formatted as (ConvNeXt / AudioProtoPNet / APEX) to compare the base classifier against the prototype-based models. Bold values highlight the best (lowest) average error rate (aEER) for each training configuration.} 
    \resizebox{\textwidth}{!}{
\begin{tabular}{@{}lcccccccc@{}} 
        \toprule
        \rule{0pt}{2.2ex}
        Training Set & \multicolumn{1}{c}{MelGAN} & \multicolumn{1}{c}{MelGAN (L)} & \multicolumn{1}{c}{MB-MelGAN} & \multicolumn{1}{c}{FB-MelGAN} & \multicolumn{1}{c}{HiFi-GAN} &\multicolumn{1}{c}{PWG} & \multicolumn{1}{c}{WaveGlow} & \multicolumn{1}{c}{\textbf{aEER}} \\ 
        \midrule 
        \rule{0pt}{2.4ex}MelGAN     & 0.0 / 0.0 / 0.0 & 0.0 / 0.0 / 0.0 & 22.9 / 27.6 / 22.9 & 40.0 / 39.7 / 40.0 & 23.4 / 28.9 / 23.4 & 10.0 / 14.2 / 10.0 & 30.4 / 34.0 / 30.4 & \textbf{18.1} / 20.6 / \textbf{18.1}  \\
        \rule{0pt}{2.4ex}MelGAN (L) & 0.0 / 0.0 / 0.0 & 0.0 / 0.0 / 0.0 & 27.1 / 28.5 / 27.1 & 39.6 / 40.5 / 39.6 & 27.2 / 27.2 / 27.2 & 15.9 / 18.0 / 15.9 & 31.0 / 29.1 / 31.0 & \textbf{20.1} / 20.5 / \textbf{20.1} \\
        \rule{0pt}{2.4ex}MB-MelGAN  & 3.2 / 1.9 / 3.2 & 3.4 / 1.9 / 3.4 & 0.0 / 0.0 / 0.0 & 13.4 / 4.2 / 13.4 & 7.9 / 3.7 / 7.9 & 2.9 / 1.0 / 2.9 & 11.0 / 3.9 / 11.0 & 6.0 / \textbf{2.4} / 6.0 \\
        \rule{0pt}{2.4ex}FB-MelGAN  & 8.3 / 4.9 / 8.3 & 0.1 / 0.2 / 0.1 & 0.4 / 0.5 / 0.4 & 0.0 / 0.0 / 0.0 & 0.1 / 0.2 / 0.1 & 3.2 / 1.8 / 3.2 & 10.1 / 5.8 / 10.1 & 3.2 / \textbf{1.9} / 3.2 \\
        \rule{0pt}{2.4ex}HiFi-GAN   & 0.0 / 0.1 / 0.0 & 0.0 / 0.0 / 0.0 & 0.2 / 0.1 / 0.2 & 1.6 / 0.7 / 1.6 & 0.0 / 0.0 / 0.0 & 1.4 / 0.5 / 1.4 & 9.6 / 2.5 / 9.6 & 1.8 / \textbf{0.6} / 1.8 \\
        \rule{0pt}{2.4ex}PWG        & 1.4 / 0.8 / 1.4 & 3.4 / 0.8 / 3.4 & 49.1 / 25.5 / 49.1 & 48.9 / 34.6 / 48.9 & 32.8 / 12.3 / 32.8 & 0.0 / 0.0 / 0.0 & 50.0 / 30.6 / 50.0 & 26.5 / \textbf{14.9} / 26.5 \\
        \rule{0pt}{2.4ex}WaveGlow   & 9.9 / 3.8 / 9.9 & 29.0 / 22.1 / 29.0 & 13.8 / 10.1 / 13.8 & 33.8 / 33.0 / 33.8 & 19.2 / 19.8 / 19.2 & 5.1 / 1.7 / 5.1 & 0.0 / 0.0 / 0.0 & 15.9 / \textbf{12.9} / 15.9 \\
        \bottomrule
    \end{tabular}}
    \label{tab:deepfake_eval}
\end{table*}

\begin{table*}[ht]
    \caption{Impact of targeted spectrogram masking on deepfake detection performance. Columns denote the specific vocoder dataset used for evaluation. Baseline model was trained on the HiFi-GAN dataset. Values indicate the Equal Error Rate (EER) [\%], across all evaluation sets, under three conditions: no masking, random masking, and \APEX{}-guided masking across the square, time, frequency, and time-frequency extraction schemes. Random masking results are averaged across four different seeds and results are presented in the $\text{avg}\pm\text{std}$ form. Random masks of each type has exactly the same same as masks created with \APEX{}.}
    \centering
    \resizebox{\textwidth}{!} {
    \begin{tabular}{l|l|c|c|c|c|c|c|c|c}
        \hline
        & & MelGAN & MelGAN (L) & MB-MelGAN &  FB-MelGAN & HiFi-GAN & PWG & WaveGlow & \textbf{aEER} \\ \hline
        No mask &  & 0.0 & 0.0 & 0.2 & 1.6 & 0.0 & 1.4 & 9.6 & 1.8 \\ \hline
        \multirow{4}{*}{\shortstack[l]{Random \\ mask}} & Square & $0.0\pm0.0$ & $0.0\pm0.0$ & $0.2\pm0.1$ & $1.6\pm0.1$ & $0.0\pm0.0$ & $1.4\pm0.1$ & $9.8\pm0.2$ & $\mathbf{1.9\pm0.1}$\\ 
        & Time & $0.0\pm0.0$ & $0.0\pm0.0$ & $0.3\pm0.1$ & $1.9\pm0.2$ & $0.0\pm0.0$ & $0.4\pm0.2$ & $9.9\pm0.2$ & $\mathbf{1.9 \pm0.1}$\\ 
        & Frequency & $0.1\pm0.1$ & $0.0\pm0.0$ & $0.4\pm0.1$ & $5.2\pm0.4$ & $0.0\pm0.0$ & $2.4\pm0.2$ & $13.8\pm0.2$ & $3.1\pm0.1$\\ 
        & Time-frequency & $0.1\pm0.1$ & $0.0\pm0.1$ & $0.7\pm0.1$ & $6.0\pm0.4$ & $0.0\pm0.0$ & $2.1\pm0.2$ & $13.8\pm0.4$ & $3.2\pm0.1$\\ \hline
        \multirow{4}{*}{\shortstack[l]{\APEX{} \\ mask}} & Square & 0.0 & 0.0 & 0.2 & 1.6 & 0.0 & 1.4 & 10.0 & \textbf{1.9}\\ 
        & Time & 0.1 & 0.0 & 0.3 & 1.9 & 0.0 & 1.5 & 9.4 &  \textbf{1.9} \\ 
        & Frequency & 0.0 & 0.0 & 0.7 & 7.8 & 0 & 2.7 & 12.7 & \textbf{3.4}\\ 
        & Time-frequency & 0.0 & 0.0 & 0.4 & 7.7 & 0.0 & 2.5 & 13.8 & \textbf{3.5} \\ \hline
    \end{tabular}}
    \label{tab:deepfake_masking}
\end{table*}

\textbf{Results} We evaluate the classification performance using the Equal Error Rate (EER) and average EER (aEER) across the seven unseen vocoder test datasets, as reported in Tab.~\ref{tab:deepfake_eval}. The results demonstrate that the \APEX{} framework successfully preserves the predictive capabilities of the underlying ConvNeXt backbone, achieving identical error rates across various training configurations. To assess the essentiality of the regions highlighted by our method, we conducted a targeted spectrogram masking ablation study. For this evaluation, we specifically chose the baseline model trained on the HiFi-GAN dataset due to its superior baseline performance demonstrated in Tab.~\ref{tab:deepfake_eval}. Tab.~\ref{tab:deepfake_masking} compares the EER under three conditions: unmasked inputs, random masking, and \APEX{}-guided masking utilizing our four extraction schemes (Square, Time, Frequency, and Time-frequency). While random masking produced only negligible increases in the error rate, masking the specific regions identified by \APEX{} resulted in a higher EER across all extraction schemes. This pronounced performance degradation confirms that the \APEX{} method accurately localizes the critical acoustic artifacts upon which the model heavily relies to distinguish between bona fide and fake audio.

Furthermore, a qualitative visual inspection of the \APEX{} explanations is provided in Fig.~\ref{fig:comp_deepfake}, which illustrates the model's behavior on correctly classified real audio from the LJSpeech dataset. The figure displays the input query spectrograms and their corresponding heatmaps alongside the extracted prototypical parts. By highlighting the spectral differences between the real audio and the audio vocoded with HiFi-GAN, Fig.~\ref{fig:comp_deepfake} qualitatively confirms that \APEX{} produces highly interpretable and semantically meaningful boundaries that align with the synthesis artifacts present in the data.

\begin{table*}[ht]
    \caption{Multi-label bioacoustic classification results on the BirdSet~\cite{birdset} test sets. The table compares cmAP, AUROC, and T1-Acc across eight geographical regions. Bold values indicate the best performance per column. Crucially, the scores for our post-hoc method precisely match those of the baseline ConvNeXt, empirically confirming that our disentanglement module maintains strict output invariance while enabling interpretability.}
    \centering
    \renewcommand{\arraystretch}{1.1} 
    \setlength{\tabcolsep}{4pt}       
    \begin{tabular}{c|c|cccccccc}
        \hline
         & & POW & PER & NES & UHH & HSN & NBP & SSW & SNE  \\ 
        \hline
        \multirow{3}{*}{ConvNeXt}
        & cmAP   & 0.43 & 0.23 & \textbf{0.38} & 0.24 & 0.52 & 0.68 & 0.41 & 0.32  \\ 
        & AUROC  & 0.85 & 0.72 & 0.89 & 0.70 & 0.89 & 0.92 & 0.93 & 0.82  \\
        & T1-Acc & 0.84 & 0.46 & \textbf{0.53} & \textbf{0.56} & \textbf{0.70} & 0.71 & 0.55 & 0.71  \\ 
        \hline
        \multirow{3}{*}{AudioProtoPNet}
        & cmAP   & \textbf{0.52} & \textbf{0.30} & 0.34 & \textbf{0.32} & \textbf{0.55} & \textbf{0.70} & \textbf{0.43} & \textbf{0.33}  \\   
        & AUROC  & \textbf{0.90} & \textbf{0.78} & \textbf{0.94} & \textbf{0.86} & \textbf{0.93} & \textbf{0.93} & \textbf{0.97} & \textbf{0.86}  \\
        & T1-Acc & \textbf{0.88} & \textbf{0.60} & 0.52 & 0.49 & 0.67 & \textbf{0.73} & \textbf{0.66} & \textbf{0.76}  \\ 
        \hline
        \multirow{3}{*}{\APEX{} (our)}
        & cmAP   & 0.43 & 0.23 & \textbf{0.38} & 0.24 & 0.52 & 0.68 & 0.41 & 0.32  \\ 
        & AUROC  & 0.85 & 0.72 & 0.89 & 0.70 & 0.89 & 0.92 & 0.93 & 0.82  \\
        & T1-Acc & 0.84 & 0.46 & \textbf{0.53} & \textbf{0.56} & \textbf{0.70} & 0.71 & 0.55 & 0.71  \\ 
        \hline
    \end{tabular}
    \label{tab:results_birdset_eval}
\end{table*}

\begin{table*}[ht]
    \caption{Evaluation of the essentiality of \APEX{}-highlighted spectrogram regions on the SNE test set from the Birdset dataset \cite{birdset}. The table compares classification metrics (cmAP, AUROC, T1-Acc) for unmasked inputs, random masking, and targeted APEX masking across square, time, frequency, and time-frequency prototype extraction schemes. A significant performance drop under APEX masking indicates the importance of these localized regions. Random masking results are averaged across four different seeds. In all cases standard deviation was less than $0.1$. \APEX{} masking results are deterministic.}
    \centering
    \begin{tabular}{l|l|c|c|c|c}
        \hline
        & & Square & Time & Frequency & Time-frequency \\ \hline
        \multirow{3}{*}{No masking} & cmAP & 0.32 & 0.32 & 0.32 & 0.32 \\ 
        & AUROC & 0.82 & 0.82 & 0.82 & 0.82 \\ 
        & T1-Acc & 0.71 & 0.71 & 0.71 & 0.71 \\ \hline
        \multirow{3}{*}{Random masking} & cmAP & 0.31 & 0.31 & 0.27& 0.27 \\ 
        & AUROC & 0.81 & 0.81 & 0.79 & 0.78 \\ 
        & T1-Acc & 0.70 & 0.70 & 0.63 & 0.62 \\ \hline
        \multirow{3}{*}{\APEX{} masking} & cmAP & \textbf{0.29} & \textbf{0.28} & \textbf{0.20} & \textbf{0.17} \\ 
        & AUROC & \textbf{0.80} & \textbf{0.79} & \textbf{0.75} & \textbf{0.73} \\ 
        & T1-Acc & \textbf{0.63} & \textbf{0.62} & \textbf{0.37} & \textbf{0.32} \\ \hline
    \end{tabular}
    \label{tab:results_birdset_masking}
\end{table*}

\subsection{Bioacoustic Classification}

\textbf{Training} We further evaluate our approach on the task of multi-label bird sound classification. Following the training procedure established by AudioProtoPNet \cite{audioprotopnet}, we first train a baseline ConvNeXt-Base model on the XCL subset of the BirdSet dataset \cite{birdset}, which encompasses 9,734 bird species and over 6,800 hours of audio recordings. While AudioProtoPNet is configured using the default settings from its original publication, the APEX model is trained on the XCM, a smaller subset of the XCL.

\textbf{Results} Performance was evaluated across the test datasets of BirdSet, covering different geographical regions, using class-mean average precision (cmAP), AUROC, and Top-1 Accuracy (T1-Acc). As detailed in Tab.~\ref{tab:results_birdset_eval}, the classification metrics for our post-hoc \APEX{} method precisely match those of the baseline ConvNeXt model across all evaluated geographical regions. This empirically confirms that our disentanglement module maintains strict output invariance, successfully enabling interpretability without altering the underlying model's predictive capabilities. In contrast, the inherently interpretable AudioProtoPNet exhibits performance trade-offs that differ from those of the base classifier. 

To quantitatively assess whether the specific spectrogram regions highlighted by \APEX{} represent essential acoustic features, we conducted a targeted masking ablation study on the SNE test set. Tab.~\ref{tab:results_birdset_masking} compares the model's performance on unmasked inputs against random masking and \APEX{}-guided masking across the square, time, frequency, and time-frequency extraction schemes. While random masking yielded only minor performance degradation (e.g., dropping the cmAP from 0.32 to 0.27), under the time-frequency scheme, targeted \APEX{} masking caused a severe performance drop, reducing the cmAP to 0.17. This significant drop across all metrics indicates that the localized regions identified by \APEX{} are highly essential for the model's decision-making process.

Finally, Fig.~\ref{fig:comp_birdset} provides a qualitative comparison between APEX and AudioProtoPNet  using a correctly classified Mountain Chickadee (mouchi) from the SNE test set. Visual inspection confirms that APEX produces highly localized and consistent time-frequency explanations. Compared to AudioProtoPNet, our method bounds the relevant acoustic events with greater semantic clarity and precision.

\section{Conclusions}

In this work, we introduced \APEX{}, a post-hoc prototype-based interpretability framework for arbitrary pre-trained audio classifiers. \APEX{} goes beyond spectrogram-attribution adaptations and prototype networks trained from scratch by operating on the model's latent representation with strict output invariance. We insert a learnable invertible transformation between backbone and classification head and compensate it in the head, thereby reorganizing the feature basis into semantically purer channels without changing the original logits.

Four prototype extraction schemes with audio-specific priors on the time-frequency plane, including square-based, time-based, frequency-based, and hybrid time-frequency prototypes are proposed to capture transient localized events, temporally defined patterns, stable spectral evidence, and balanced cues, grounding explanations in interpretable acoustic structure rather than generic activations to reduce ambiguity in acoustic similarity. Experiments on WaveFake deepfake detection and BirdSet large-scale bioacoustic classification show that \APEX{} preserves baseline performance exactly while producing well-localized, semantically meaningful explanations. In deepfake detection, Equal Error Rates match the underlying ConvNeXt across vocoder-specific training regimes, and masking \APEX{}-identified regions increases error more than random masking, indicating localization of evidence used by the classifier. In BirdSet, \APEX{} scales to thousands of classes while matching baseline cmAP, AUROC, and Top-1 accuracy across regions, and \APEX{}-guided masking yields large drops in these metrics. Qualitative results indicate clearer and more localized time-frequency explanations than Grad-CAM and LIME, and more precise localization than existing prototype-based methods.

Future work will extend \APEX{} to self-supervised and foundation models such as Wav2Vec2.0~\cite{wav2vec2.0} and HuBERT~\cite{hubert} that use contextualized waveform embeddings and exhibit hierarchical, layer-dependent latent structure. Key directions include adapting the invertible transformation to transformer encoders by layer-wise disentanglement and cross-layer prototype consistency to study where stable acoustic concepts emerge, and developing quantitative evaluation of faithfulness, stability, and robustness under time stretching, pitch shifting, and additive noise. We also plan to study multimodal and generative settings to test if structurally grounded prototypes provide consistent explanations fo discriminative and generative models.

\textbf{Limitations} \APEX{} is applicable only to architectures whose classification head comprises a pooling layer above the backbone followed by a single-layer classifier; these architectures operate on the spectrogram inputs.

\newpage
\bibliographystyle{IEEEtran}
\bibliography{mybib}

\end{document}